\documentclass[11pt,twoside,a4paper,fleqn,titlepage]{article}
\pdfoutput=1
\usepackage[english]{babel}
\usepackage{graphicx}
\usepackage{subfigure}
\usepackage{styles/cite}
\usepackage{hyperref}
\usepackage{amsmath,amsfonts,amssymb,mathrsfs}
\makeatletter 
\long\def\@makecaption#1#2{\footnotesize
\vskip\abovecaptionskip 
\sbox\@tempboxa{#1: #2}
\ifdim \wd\@tempboxa >\hsize
#1: #2\par \else
\global \@minipagefalse
\hb@xt@\hsize{\hfil\box\@tempboxa\hfil}
\fi
\vskip\belowcaptionskip} 
\makeatother
\def\SU2U1{{\rm SU}(2)\times{\rm U}(1)}

\mathchardef\qsm=63
\mathchardef\pls=43
\mathchardef\mns=512
\mathchardef\plm=518
\mathchardef\eql=61
\mathchardef\smallleft=300
\mathchardef\smallright=301
\mathchardef\perslsh=47
\mathchardef\les=316
\mathchardef\gre=318
\mathchardef\leq=532
\mathchardef\grq=533
\chardef\usc=95
\chardef\til=126

\def\sqr#1#2#3{{\vcenter{\hrule height.#3ex\hbox{\vrule width.#2ex height#1ex
    \kern#1ex\vrule width.#3ex}\hrule height.#2ex}}}

\def\angleto{\vrule width.035em height2.1ex depth-.56ex\unskip\kern-.6ex\to}
\def\perchc#1{{\raise.4ex\hbox{$\mkern4mu#1{\it\perslsh}_
             {\mkern-5mu\scriptscriptstyle{{\rm o}\!{\rm o}}}^
             {\mkern-12.8mu\scriptscriptstyle{\rm o}}$}}}

\catcode`\@=11 
\def\parenbar{\mathpalette\p@renb@r}
\def\p@renb@r#1#2{\vbox{%
  \ifx#1\scriptscriptstyle \dimen@.7em\dimen@ii.2em\else
  \ifx#1\scriptstyle \dimen@.8em\dimen@ii.25em\else
  \dimen@1em\dimen@ii.4em\fi\fi \offinterlineskip
  \ialign{\hfill##\hfill\cr
    \vbox{\hrule width\dimen@ii}\cr
    \noalign{\vskip-.3ex}%
    \hbox to\dimen@{$\mathchar300\hfil\mathchar301$}\cr
    \noalign{\vskip-.3ex}%
    $#1#2$\cr}}}
\catcode`\@=12 

\newbox\struttbox
\setbox\struttbox=\hbox{\vrule height1.65ex depth.485ex width0pt}
\def\strutt{\relax\ifmmode\copy\struttbox\else\unhcopy\struttbox\fi}
\def\stru#1#2{\relax\ifmmode\hbox{\vrule height#1 depth#2 width0pt}
\else\vrule height#1 depth#2 width0pt\fi}

\def\ronum#1{\uppercase\expandafter{\romannumeral#1}}
\def\ronuml#1{\expandafter{\romannumeral#1}}
 {\end{list}}
 {\end{list}}
 {\end{list}}

\catcode`\@=11 
\newlength{\@fninsert}
\newlength{\@fnwidth}
\renewcommand{\@makefntext}[1]%
  {\noindent\makebox[\@fninsert][r]{\@makefnmark}\hfil%
  \parbox[t]{\@fnwidth}{#1}}
\catcode`\@=12 
\catcode`\@=11 
\renewcommand\section{\@startsection{section}{1}{\z@}%
                                   {-3.5ex \@plus -1ex \@minus -.2ex}%
                                   {2.3ex \@plus.2ex}%
                                   {\normalfont\Large\bfseries}}
\renewcommand\subsection{\@startsection{subsection}{2}{\z@}%
                                   {-3.25ex\@plus -1ex \@minus -.2ex}%
                                   {1.5ex \@plus .2ex}%
                                   {\normalfont\large\bfseries}}
\renewcommand\subsubsection{\@startsection{subsubsection}{3}{\z@}%
                                   {-3.25ex\@plus -1ex \@minus -.2ex}%
                                   {1.5ex \@plus .2ex}%
                                   {\normalfont\large\bfseries}}
\renewcommand\paragraph{\@startsection{paragraph}{4}{\z@}%
                                   {3.25ex \@plus1ex \@minus.2ex}%
                                   {1.2ex \@plus .2ex}%
                                   {\normalfont\normalsize\bfseries}}
\catcode`\@=12 

\def\drftdate{17th May 2010}
\begin{document}
%
%
\pagestyle{plain}
\setcounter{secnumdepth}{5}
\setcounter{tocdepth}{2}
\selectlanguage{english}
\makeatletter
\renewcommand{\l@subsection}{\@dottedtocline{2}{1.0em}{2.3em}}
\renewcommand{\l@subsubsection}{\@dottedtocline{3}{2.0em}{3.2em}}
\renewcommand{\l@paragraph}{\@dottedtocline{4}{3.0em}{4.1em}}
\renewcommand{\l@subparagraph}{\@dottedtocline{5}{4.0em}{5em}}
\makeatother
%
%
\pagestyle{empty}
\begin{figure}
  \vspace{-1cm}
  \begin{center}
    \includegraphics[width=\textwidth]{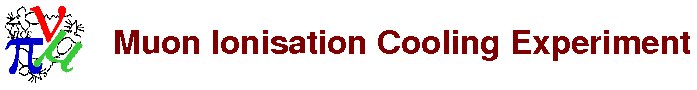}
  \end{center}
  \includegraphics[width=0.25\textwidth]{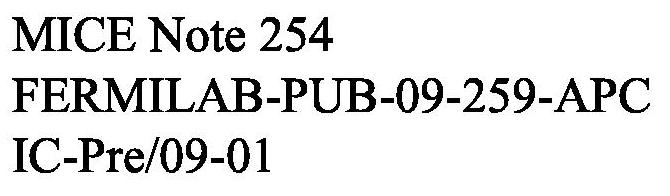}\\
  \rightline{\drftdate}
  \vspace{1.5cm}
\end{figure}
\begin{center}
  {\bf\LARGE
    The design, construction and performance of the \\ 
	\vspace{5mm}
    MICE scintillating fibre trackers
  }
\end{center}
%
%
\vspace{1.5cm}
\section*{Abstract}

Charged-particle tracking in the international Muon Ionisation
Cooling Experiment (MICE) will be performed using two solenoidal
spectrometers, each instrumented with a tracking detector based on 
350~$\mu$m diameter scintillating fibres.
The design and construction of the trackers is described along with 
the quality-assurance procedures, photon-detection system, readout 
electronics, reconstruction and simulation software and the data-acquisition system. 
Finally, the performance of the MICE tracker, determined using cosmic
rays, is presented.


%
\clearpage
%
%
{\noindent
  M.~Ellis, P.R.~Hobson, P.~Kyberd, J.J.~Nebrensky
  \\{\it
    Brunel University, Uxbridge, Middlesex, UB8 3PH, UK
  } 
  \par \filbreak
  A.~Bross, J.~Fagan, T.~Fitzpatrick, R.~Flores, R.~Kubinski, 
  J.~Krider, R.~Rucinski, P.~Rubinov, C.~Tolian
  \\{\it
    Fermilab, P.O. Box 500, Batavia, IL 60510-0500, USA
  } 
  \par \filbreak
  T.L.~Hart\footnote{Now at University of Mississippi, Oxford, 
                   MS 38677, USA}, 
  D.M.~Kaplan, W.~Luebke, B.~Freemire, M.~Wojcik
  \\{\it
    Physics Division, Illinois Institute of Technology, 3101 S. Dearborn St., Chicago, IL 60616, 
    USA
  } 
  \par \filbreak
  G.~Barber, D.~Clark, I.~Clark, P.J.~Dornan, A.~Fish, S.~Greenwood,
  R.~Hare, A.~Jamdagni, V.~Kasey, M.~Khaleeq, J.~Leaver, K.R.~Long,
  E.~McKigney\footnote{Now at Los Alamos Natl. Lab., P.O. Box 1663,
                       Los Alamos, NM 87545, USA}, 
  T.~Matsushita\footnote{Now at Kobe University, Faculty of Science,
                         1-1 Rokkodai-cho, Nada-ku, Kobe-shi, 
                         Hyogo 657-8501, Japan},
  C.~Rogers\footnote{Now at STFC Rutherford Appleton Laboratory, Chilton, 
                     Didcot, Oxon, OX11 0QX, UK},
  T.~Sashalmi, P.~Savage, 
  M.~Takahashi\footnote{Now at the University of Manchester,
                        School of Physics and Astronomy, Schuster 
                        Laboratory, Manchester M13 9PL, UK},
  A.~Tapper
  \\{\it
    Department of Physics, Blackett Laboratory, Imperial College
    London, Exhibition Road, London SW7 2AZ, UK
  } 
  \par \filbreak
  K.~Yoshimura
  \\{\it
    High Energy Accelerator Research Organization (KEK), 
    Institute of Particle and Nuclear Studies, Tsukuba, 
    Ibaraki, Japan
  } 
  \par \filbreak
  P.~Cooke, R.~Gamet
  \\{\it
    Department of Physics, Oliver Lodge Laboratory, 
    University of Liverpool, Liverpool, L69 7ZE, UK
  } 
  \par \filbreak
  H.~Sakamoto, Y.~Kuno, A.~Sato, T.~Yano, M.~Yoshida
  \\{\it
    Graduate School of Science, Department of Physics, 
    Osaka University, Toyonaka, Osaka, Japan
  } 
  \par \filbreak
  C.~MacWaters
  \\{\it
    STFC Rutherford Appleton Laboratory, Chilton, Didcot, 
    Oxfordshire, OX11 0QX, UK
  } 
  \par \filbreak
  L.~Coney, G.~Hanson,  A.~Klier\footnote{Now at Weizmann Institute of Science, Department of Particle Physics, 
P.O. Box 26, Rehovot 76100, Israel}
  \\{\it
    University of California, Riverside, Riverside, CA 92521-0413 USA
  } 
  \par \filbreak
  D.~Cline, X.~Yang
  \\{\it
    University of California at Los Angeles Physics Department, 
    Los Angeles, CA 90024, USA
  } 
  \par \filbreak
  D.~Adey
  \\{\it
    Department of Physics, University of Warwick, Coventry,
    CV4 7AL, UK
  } 
  \par \filbreak
}

\clearpage
%
%
\pagestyle{plain}
\pagenumbering{roman}                   
\setcounter{page}{1}
%
%
\tableofcontents
\clearpage
%
%
\pagestyle{plain}
\parindent 10pt
\pagenumbering{arabic}                   
\setcounter{page}{1}
%
%
%
\section{Introduction}
\label{Sect:Intro}

Muon storage rings have been proposed for use as sources of intense
high-energy neutrino beams in a Neutrino Factory \cite{Geer:1997iz}
and as the basis for multi-TeV lepton-antilepton colliding-beam
facilities \cite{MC}.
To optimise the performance of such facilities requires the
phase-space compression (cooling) of the muon beam prior to
acceleration and storage.  
The short muon-lifetime makes it impossible to employ traditional
techniques to cool the beam while maintaining the muon-beam intensity. 
Ionisation cooling, a process in which the muon beam is
passed through a series of liquid hydrogen absorbers interspersed with
accelerating RF cavities, is the technique proposed to cool the muon
beam. 
The international Muon Ionisation Cooling Experiment (MICE) 
will provide an engineering demonstration of the ionisation-cooling
technique and will allow the factors affecting the performance of
ionisation-cooling channels to be investigated in detail \cite{MICE}.
Muon beams of momenta between 140~MeV/c and 240~MeV/c, with
normalised emittances between 2~$\pi$mm and 10~$\pi$mm, will be
provided by a purpose-built beam line on the 800~MeV proton
synchrotron, ISIS \cite{ISIS}, at the Rutherford Appleton Laboratory
\cite{RAL}.

\begin{figure}
  \begin{center}
    \includegraphics[width=0.95\textwidth]
      {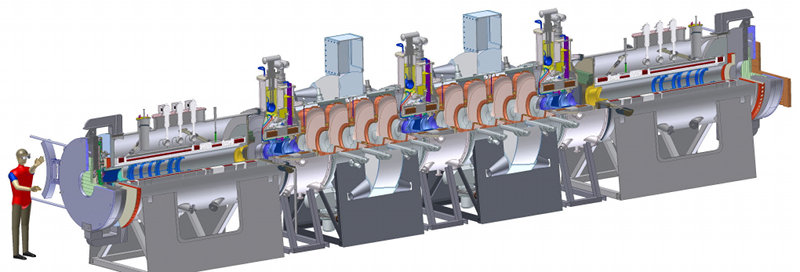}
  \end{center}
  \caption{
    Cutaway 3D rendering of the international Muon Ionisation 
    Cooling Experiment (MICE).
    The muon beam enters from the bottom left of the figure.
    The upstream PID instrumentation (not shown) is composed of two 
    time-of-flight hodoscopes (TOF0 and TOF1) and two threshold Cherenkov
    counters (CKOVa and CKOVb). The upstream spectrometer is followed by
    the MICE cooling channel, which is composed of three 20~$l$ volumes of
    liquid hydrogen and two sets of four 201~MHz accelerating cavities
    embedded in a solenoidal transport channel. This in turn is followed by
    the downstream spectrometer, a third time-of-flight hodoscope (TOF2), 
    and a calorimeter system (KL and EMR).
  }
  \label{Fig:Intro:MICE}
\end{figure}

MICE is a single-particle experiment in which the position and
momentum of each muon is measured before it enters the MICE cooling
channel and once again after it has left (see figure
\ref{Fig:Intro:MICE}) \cite{TRD}.
The MICE cooling channel, which is based on one lattice cell of the
cooling channel described in \cite{StudyII}, comprises three 20~$l$
volumes of liquid hydrogen and two sets of four 201~MHz accelerating
cavities.
Beam transport is achieved by means of a series of superconducting
solenoids.
A particle-identification (PID) system (scintillator time-of-flight
hodoscopes TOF0 and TOF1 and threshold Cherenkov counters CKOVa and
CKOVb) upstream of the cooling channel allows a pure muon beam to be
selected. 
Downstream of the cooling channel, a final hodoscope (TOF2) and a
calorimeter system allow muon decays to be identified.  
The calorimeter is composed of a KLOE-like lead-scintillator section (KL)
followed by a fully active scintillator detector (the electron-muon ranger, 
EMR) in which the muons are brought to rest. 
For a full description of the experiment see \cite{TRD}.

Charged-particle tracking in MICE is provided by two solenoidal
spectrometers.
Together, the spectrometers are required to determine the expected relative
change in transverse emittance of approximately 10\% with a precision
of $\pm 1\%$ (i.e. a 0.1\% measurement of the absolute emittance).  
The trackers themselves are required to have high track-finding
efficiency in the presence of background induced by X-rays produced in
the RF cavities. 
Each spectrometer consists of a 4~T superconducting solenoid of 40~cm
bore instrumented with a tracker composed of five planar
scintillating-fibre stations.  
Each station is composed of three doublet-layers of scintillating
fibres laid out in a `$u,v,w$' arrangement.  
To reduce multiple Coulomb scattering of muons to an acceptable level,
a fibre diameter of 350~$\mu$m is required. 
The scintillation light is read out via 1.05~mm clear-fibre
light-guides. 
To reduce the cost of the readout electronics seven 350~$\mu$m fibres
are read out through each clear-fibre light-guide.
The active area of each station is a circle of diameter 30~cm.

The concentration of the primary and secondary dopants within the
scintillating fibre must be chosen to maximise the light yield
while minimising the fibre-to-fibre optical cross talk. 
The passage of a charged particle through the fibre causes energy to
be transferred to the primary dopant, para-terphenyl (pT). 
The peak of the scintillation light spectrum of pT is at a wavelength
of $\sim 350$~nm \cite{pTlight}.
The secondary dopant, 3-hydroxflavone (3HF), absorbs this light and
re-emits it at a wavelength of $\sim 525$~nm \cite{3HFlight}. 
The concentration of primary dopant must be such that efficient energy
transfer from the polymer to the pT occurs.  
The energy-transfer process, known as the F\"orster Transfer
\cite{Forster} process, occurs when the mean distance between a
polymer molecule and a pT molecule is on the order of 10 angstroms.  
It is this criterion that determines the minimum concentration
required for efficient transfer. 
The concentration of 3HF must be small enough to ensure negligible
secondary light attenuation along the length of the active fibre, but
large enough that the absorption length of the primary light in the
3HF is small compared to the fibre diameter.
The latter condition ensures that fibre-to-fibre cross talk is
eliminated.
Light-yield measurements using prototype stations led to the choice of
1.25\% and 0.25\% by weight for the concentrations of pT and 3HF
respectively \cite{Doping}. 
Light with wavelength shorter than around 450~nm damages the secondary
dopant (3HF) and causes discolouration of the fibres and therefore
reduces the light yield.
For this reason, filtered light was used throughout the fabrication process
and steps in the process were photographed without the use of flash.

The MICE trackers are read out using the D\O{}{} Central Fiber Tracker
(CFT) optical readout and electronics system \cite{D0NIM}.
The scintillation light is detected using Visible Light Photon
Counters (VLPCs) \cite{VLPC,VLPC1}.
These are low band-gap silicon avalanche devices that are
operated at 9~K. 
The VLPCs have a high quantum efficiency ($\sim 80\%$) and a high gain
that in some devices is in excess of $50\,000$. 
The VLPC signals are digitised using the Analogue Front End with
Timing (AFE~IIt) board developed by the D\O{} collaboration
\cite{AFEIIt}.

The solenoidal field in the tracking volume is designed to be uniform at the 3 per
mil level. 
The MICE coordinate system is such that the $z$ axis is parallel to
the beam, the $y$ axis points vertically upwards, and the $x$ axis
completes a right-handed coordinate system.
A muon therefore describes a circle in the $x,y$ plane as it travels
through the solenoid. 
The transverse momentum of the muon is obtained by determining the
radius of this circle, while the number of turns determines the
$z$-component.
The station spacing has been chosen to optimise the performance of the
reconstruction (track-finding efficiency and parameter resolution).

This paper is organised as follows.
The mechanical design and construction of the trackers are described
in section \ref{Sect:Mechanical}.
The photon-detection system and readout electronics are presented in
sections \ref{Sect:PhotonDetection} and \ref{Sect:Electronics}
respectively.
Section \ref{Sect:Performance} contains a summary of the performance
of the devices.
Finally, a summary is presented in section \ref{Sect:Summary}.

%
\section{Mechanical design and construction}
\label{Sect:Mechanical}

%

\begin{figure}
  \begin{center}
    \includegraphics[width=\textwidth]%
    {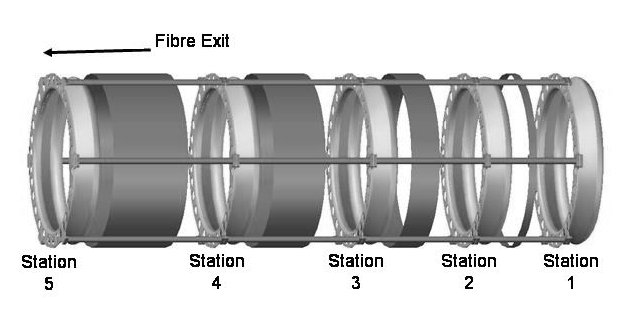}
  \end{center}
  \caption{
    Schematic diagram of the MICE tracker.
    The five stations are shown supported by the carbon-fibre 
    space frame, with fibres omitted for clarity.
    The station numbering scheme is indicated together with the
    direction in which the clear-fibre light-guides leave the tracking
    volume.
  }
  \label{Fig:MechIntro:Layout}
\end{figure}

The layout of the MICE tracker is shown in
figure \ref{Fig:MechIntro:Layout}.
The five stations are held in position using a carbon-fibre
space-frame.
The distance between neighbouring stations is such that each
nearest-neighbour spacing is unique.
This ensures that the azimuthal rotation of track position from one
station to the next differs, this difference being important in
resolving ambiguities at the pattern-recognition stage.  
The station spacing, together with other key parameters of the tracker
module, are presented in table \ref{Tab:MechIntro:Parameters}.
\begin{table}
  \caption{
    Key parameters of the MICE tracker module.
    The first section of the table presents the parameters of the
    scintillating-fibre tracker itself.
    The second section reports the environment within the tracking
    volume in which the tracker must operate.
    The final section reports the parameters of the MICE spectrometer
    solenoid that directly affect the performance of the tracker.
  }
  \label{Tab:MechIntro:Parameters}
  \begin{center}
    \begin{tabular}{|l|l|l|}
      \hline
      {\bf Component}                   & {\bf Parameter}                 & {\bf Value    }           \\
      \hline\hline
      {\bf Scintillating fibre} & Scintillating fibre diameter            & 350~$\mu$m                \\
      {\bf tracker}             & Primary dopant, pT, concentration       & 1.25\% (by weight)        \\
                                & Secondary dopant, 3HF, concentration    & 0.25\% (by weight)        \\
                                & Fibre pitch                             & 427~$\mu$m                \\
                                & Estimated light yield per single fibre  & 10 photo-electrons         \\
                                & Number of scintillating fibres per      &                          \\
                                & optical readout channel                 &  7                          \\
                                & Position resolution per plane           & 470~$\mu$m                \\
                                & Views per station                       & 3                         \\
                                & Radiation length per station            & 0.45\%		      \\
                                & Stations per spectrometer               & 5                         \\
                                & Station separation: 1 - 2               & 20 cm                     \\
                                & Station separation: 2 - 3               & 25 cm                     \\
                                & Station separation: 3 - 4               & 30 cm                     \\
                                & Station separation: 4 - 5               & 35 cm                     \\
      \hline
      {\bf Tracking volume}     & Sensitive volume: length                & 110 cm                    \\
                                & Sensitive volume: diameter              & 30 cm                     \\
                                & Gas in warm bore of spectrometer        & Helium at atmospheric     \\
                                & solenoid                                &  pressure                 \\
      \hline
      {\bf Spectrometer}        & Magnetic field in tracking volume       & 4 T                       \\
      {\bf solenoid}            & Field uniformity in tracking volume     & 3 per mil                 \\
                                & Field stability                         & 1 per mil                 \\
                                & Warm bore diameter                      & 40 cm                     \\
      \hline
    \end{tabular}
  \end{center}
\end{table}

\begin{figure}
  \begin{center}
    \includegraphics[width=0.8\textwidth]
    {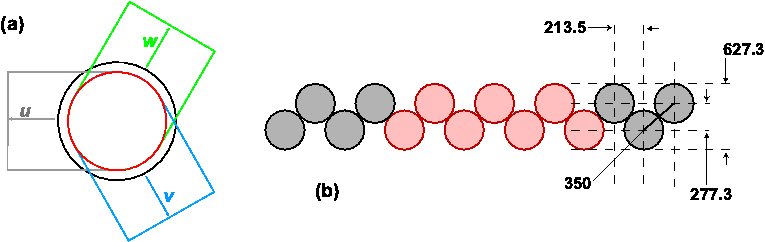}
  \end{center}
  \caption{
    (a) Arrangment of the doublet layers in the scintillating-fibre
    stations.
    The outer circle shows the solenoid bore while the inner circle
    shows the limit of the active area of the tracker.
    The grey, blue, and green arrows indicate the direction that
    individual 350~$\mu$m fibres run in the $u, v$, and $w$ planes
    respectively.
    (b) Detail of the arrangement of the scintillating fibres in a
    doublet layer.
    The fibre spacing and the fibre pitch are indicated on the
    right-hand end of the figure in $\mu$m.  
    The pattern of seven fibres ganged for readout in a single
    clear-fibre light-guide is shown in red.
  }
  \label{Fig:MechIntro:Station}
\end{figure}

Each station consists of three `doublet layers' of 350~$\mu$m
scintillating fibres glued on a carbon-fibre station body.
The doublet layers are arranged such that the fibres in one layer run
at an angle of 120$^\circ$ to the fibres in each of the other layers
as shown in figure \ref{Fig:MechIntro:Station}a.
The arrangement of the fibres within a doublet layer is shown in
figure \ref{Fig:MechIntro:Station}b.
This packing arrangement ensures that there are no inactive regions
between adjacent fibres.
The configuration of the seven fibres ganged for readout via a single
clear-fibre light-guide is also indicated in figure
\ref{Fig:MechIntro:Station}b. 

The performance of the tracker is determined by the light yield and
the fibre pitch.
The seven-fold ganging (see figure \ref{Fig:MechIntro:Station}b),
combined with a fibre pitch of 427~$\mu$m (see table
\ref{Tab:MechIntro:Parameters}), yields an expected spatial resolution
per doublet layer of 470~$\mu$m.
The expected light yield was estimated by extrapolating that obtained
in the D\O{} fibre tracker which used 835~$\mu$m scintillating fibres
with similar dopant concentrations \cite{D0NIM}.
Taking into account the scintillating-fibre diameter used in MICE
(350~$\mu$m) and assuming a maximum clear-fibre light-guide length of
4~m, a light yield of $\sim 10$ photo-electrons is
obtained \cite{Ref:MICENote135}.

%
\subsection{The scintillating-fibre stations}
\label{SubSect:SciFiStat}

\subsubsection{Doublet-layer design and fabrication}
\label{SubSubSect:Doublets}

\begin{figure}
  \begin{center}
    \includegraphics[width=0.95\textwidth]%
    {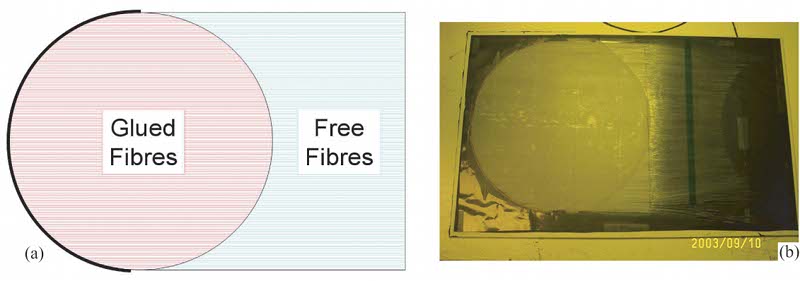}
  \end{center}
  \caption{
    (a) Schematic diagram of the scintillating-fibre ribbon.
    The circular area of diameter 32~cm that is glued to form the
    doublet layer is indicated by the red hatching.
    The mirrored end of the fibre is indicated by the solid black line.
    (b) A photograph of a completed ribbon on its substrate.
  }
  \label{Fig:SciFiStat:Ribbon}
\end{figure}

The design of the doublet layer is shown in figure
\ref{Fig:SciFiStat:Ribbon}a.  
A photograph of a completed doublet layer is shown in figure
\ref{Fig:SciFiStat:Ribbon}b.
At this stage in the manufacturing process, the scintillating-fibre
planes were referred to as `ribbons' in view of the length of
scintillating fibre extending beyond the active area.
The diameter of the circular part of the ribbon, which forms the
doublet layer, was 32~cm.
This allowed sufficient material for the ribbon to be bonded to the
station body.

The scintillating fibre used to make the ribbons was procured in the
form of 1~m long canes.
The canes were cut to length and then polished so that an aluminium mirror
could be applied to one end of the fibre. This was achieved by placing
batches of 500 canes in a water-tight former.  
Water was poured into the former and frozen so that the fibres were
supported in a block of ice.
The ice/scintillating-fibre structure was then immersed in liquid
nitrogen to ensure that the ice was hard.
Once cold, the ice/scintillating-fibre structure was polished using a
diamond tool.
After diamond polishing, the batches of fibre were thawed and an
aluminium mirror formed by vapour deposition on the polished end. 

\begin{figure}
  \begin{center}
    \includegraphics[width=0.75\textwidth]%
    {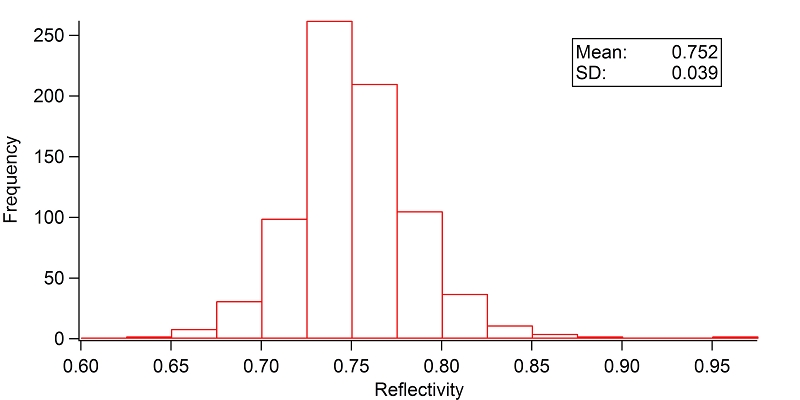}
  \end{center}
  \caption{
    Distribution of reflectivity obtained in the QA measurements
    on the aluminium mirror formed by vapour deposition on the 
    350~$\mu$m scintillating fibres used in the construction of the
    doublet layers.
    The mean and standard deviation (SD) of the distribution of
    reflectivities are shown in the inset.
  }
  \label{Fig:SciFiStat:Reflect}
\end{figure}

A sample of fibres was taken from each of the batches and the
reflectivity of the mirrors was determined as follows: a 1 m length of
fibre with a mirror on one end was excited using a near-UV LED that
excited the 3HF dopant directly.
The excitation light was detected with a photo-diode. 
The output current was measured in this configuration and then the end
of the fibre with the mirror was cut off (approximately 5 mm of fibre 
was removed). 
This end was then painted black (to kill any possible reflection from
this end) and then the LED test was repeated.
The ratio of the signal with the mirror to that without mirror
is equal to 1 + R, where R is the relectivity of the mirror.
The results of the measurements are shown in figure
\ref{Fig:SciFiStat:Reflect}.
The average reflectivity of the mirrors used in the MICE trackers is
75\% and the standard deviation of the reflectivity distribution is
approximately 4\%.

Scintillating-fibre ribbons were made following the technique
developed for the D\O{} fibre tracker \cite{D0NIM}. 
A grooved mould was machined in Delrin.
The mould was measured on a coordinate measuring machine and the mean
groove pitch was determined to be 427~$\mu$m.
Holes were drilled into the grooves to allow air to be pumped away
from the spaces within the grooves.
A Teflon release film, of thickness 25~$\mu$m, was pressed into the
mould with the aid of a vacuum established via the holes mentioned
above. 
A `tack' adhesive was then sprayed on the Teflon and the first layer
of scintillating fibres was placed in the grooves.
A circular stop, machined from a plastic sheet, was placed over the
mould so that, by butting the mirrored end of the fibre canes against
the stop, a ribbon with the proper circular active area could be
formed.
After the first layer of fibre was in the mould, spray adhesive was
applied to the fibre and the second layer of fibre (forming the
doublet layer) was placed on top of the first layer, offset by half a
fibre width. 
A polyurethane adhesive was then spread over the fibres and a
25~$\mu$m Mylar film was placed over the whole assembly.  
The assembly was then clamped to apply a uniform pressure over the
doublet layer for a period of at least 12 hours to allow the adhesive
to cure. 
The resultant `ribbon' was removed from the mould with the release
film still attached.
The final step in the ribbon fabrication was to remove carefully the
release film from the ribbon.

\subsubsection{Station body}
\label{SubSubSect:StationBody}

\begin{figure}
  \begin{center}
    \includegraphics[width=0.95\textwidth]%
    {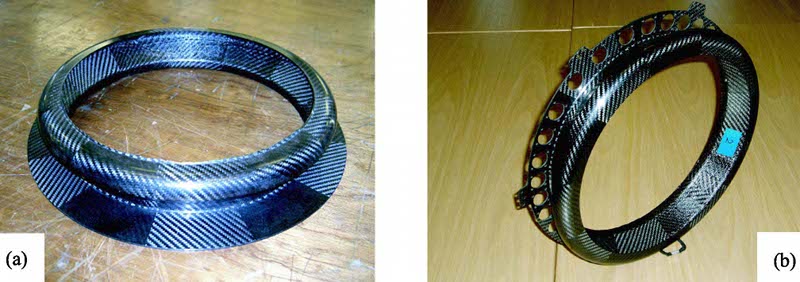}
  \end{center}
  \caption{
    (a) Photograph of a carbon fibre station body after curing in the
    autoclave. 
    (b) Photograph of the station body after final machining.  
    The holes that take the station connectors and the flanges that
    mate to the space frame are shown. 
  }
  \label{Fig:SciFiStat:Body}
\end{figure}

The station body was fabricated in carbon fibre `prepreg'
(pre-impregnated with resin) supplied by the Advanced Composites
Group (UK). 
The fibre specification was CF1300, 150~g/m$^2$, $2 \times 2$ Twill
Weave 1K HS and was impregnated with $45\%$ by weight VTM264 resin.
The station body was produced in the form of a low-mass shell capable
of supporting the scintillating fibres, with a flange for optical
connectors and for attachment to the space frame (see figure
\ref{Fig:SciFiStat:Body}a). 
The external diameter of the flange is 40~cm with the flat
scintillating-fibre contact area having an inner diameter of 30~cm and
an outer diameter of 32~cm.
A radius was formed to ensure that the fibres would not be bent by
more than the minimum bend radius as they passed from the active
surface to the optical connectors on the flange. 
The overall depth of the station is 65~mm.

A two-part female mould tool was manufactured in aluminium to the
required shape. 
The tool was doweled and bolted together to ensure accuracy. 
The surface was polished to a high gloss and degreased.
A release coating of wax was applied to the mould each time it was
used.
Due to the complex nature of the shape and the difficulty of laying-up
the carbon fibre, it was necessary to divide the station body into 12
identical $30^o$ azimuthal segments. 
A template was produced and enough parts were cut to provide three
overlapping layers (36 pieces). 
The first layer was built up on the tool from 12 segments.
Each subsequent layer was then layed with the segments offset
azimuthally by $10^o$ from the previous layer. 
To build up the flange area in preparation for a subsequent machining
process, 7 ring-shaped layers were also applied to the tool.

Once all the carbon fibre pieces were present in the layup, the 
assembly had layers of release and breather film applied before being
sealed in a vacuum bag and placed in an autoclave for cure.  
The cure cycle used was as follows: 
\begin{itemize}
  \item Apply vacuum to the inside of the bag;
  \item Raise the temperature to $80^o$~C at a rate of $0.5^o$~C/minute
        and increase the pressure on the outside of the bag to 90~psi
        at a rate of 10~psi/minute;
  \item Hold at $80^o$~C and 90~psi for four hours; and
  \item Release the external pressure and cool to $20^o$~C before
        removing the vacuum. 
\end{itemize}
Once cured, the stage was removed from the tool, dressed and made
ready for machining. 
After machining, the approximate final cured thicknesses were: fibre
plane and curved sections, 0.55~mm; flange area, 2~mm.
The station was then machined on a CNC milling machine. 
A finished station is shown in figure \ref{Fig:SciFiStat:Body}b.

\subsubsection{Station assembly}
\label{SubSubSect:StationAssembly}

\begin{figure}
  \begin{center}
    \includegraphics[width=0.95\textwidth]%
    {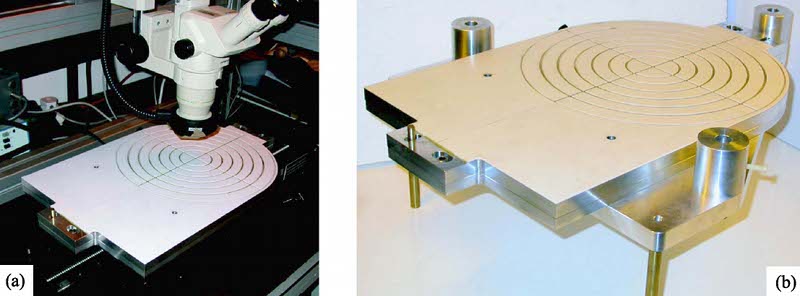}
  \end{center}
  \caption{
    (a) Photograph of the vacuum chuck in position on the linear
    stage.
    The circular and diagonal grooves cut in the white Teflon sheet
    distribute the vacuum across the doublet layer.
    The photograph also shows the microscope head; and 
    (b) Photograph of the vacuum chuck in position on the gluing jig.
  }
  \label{Fig:SciFiStat:Jigging}
\end{figure}

\begin{figure}
  \begin{center}
    \includegraphics[width=0.95\textwidth]%
    {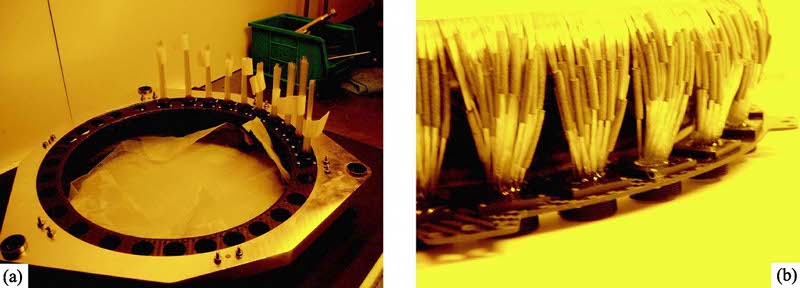}
  \end{center}
  \caption{
    (a) Photograph of station clamped in station holder.  
    The first doublet layer has been glued to the station body and the
    connectors have been fitted to the flange.
    Adhesive will now be applied to the connector face as described in
    the text.
    (b) Photograph showing in detail the fibre run from the doublet
    layer to the station connector.
    The station connectors (facing down) have been polished.
  }
  \label{Fig:SciFiStat:Potting}
\end{figure}

The first step in processing the fibre ribbons was to group the
individual fibres into the correct seven-fibre bundles. 
The bundles of seven were held together using black rubber tubes
and placed in a comb to allow the first stage of the optical
quality-assurance (QA) process to be carried out, as described in the
next section.
The bundles were then fitted into the station connectors.
The rubber tubes helped keep the fibres together as they were inserted
into the holes in the connector.
The rubber tubes, which can be seen in figure
\ref{Fig:SciFiStat:Potting}, remain on the fibre bundle to add
strength and to protect the bundle.
When all fibre bundles had been fitted to station connectors a second
optical QA procedure was carried out to ensure that the
fibres were fitted to the correct hole in the correct connector.
Once this had been accomplished, the doublet layers with connectors
fitted were aligned onto the vacuum chuck with a precision of
$\pm \sim 1$~mm and secured by vacuum.
Once the doublet layer had been secured, the chuck was aligned
precisely with respect to three locating dowels using a microscope and
linear stage (see figure \ref{Fig:SciFiStat:Jigging}a). 
 		
\begin{figure}
  \begin{center}
    \includegraphics[width=0.45\textwidth]%
    {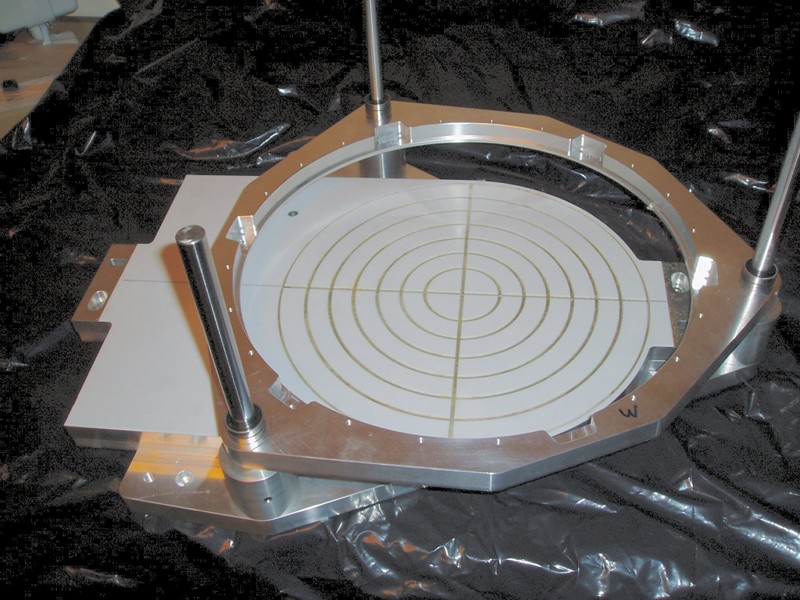}
  \end{center}
  \caption{
    Station holder positioned over vacuum chuck.
    The photograph shows the station holder in position, located by the
    three precision dowels located at the vertices of an equilateral
    triangle.
  }
  \label{Fig:SciFiStat:Jigging2}
\end{figure}

When the alignment was complete, the chuck was locked in position and
the dowels were used to locate the vacuum chuck, complete with doublet
layer, onto the assembly jig (see figure \ref{Fig:SciFiStat:Jigging}b).
The assembly jig had two main parts, the vacuum chuck holder, or
base, and the station holder; both of which are shown in figure
\ref{Fig:SciFiStat:Jigging2}. 
The station holder, as the name implies, fixed the carbon-fibre
station body in the desired location (using dowels) in relation to
the three guide shafts.
Once the station body was in the holder it remained there until all
three doublet layers had been glued in place. 
The guide shafts were set at $120^\circ$ azimuthal spacing to give the
required azimuthal alignment from doublet layer to doublet layer.

The first doublet layer was glued directly onto the carbon-fibre
station body.
Once the glue had cured, securing the plane to the station body, the 
connectors were fitted into their correct positions on the
station-body flange.
The next doublet layer was laid up on the vacuum chuck, aligned as
described above, and glued to the first doublet layer.
Once the optical connectors serving the second doublet layer had been
secured in position, the third doublet layer was glued to the second
using the same procedures.
Great care was taken to avoid producing trapped air pockets between
doublet layers since any such pockets would result in changes in the
distance between doublet layers in response to changes in barometric
pressure. 
To accomplish this, the glue was not allowed to form a complete circle
around the active area. 

When all connectors had been attached to the station body, the fibres
were potted into the connectors. 
To ensure the best possible `lay', the fibres were gently pulled and
eased into position, see figure \ref{Fig:SciFiStat:Potting}. 
When the fibres were at the best possible position, without any undue
strain, they were potted.
To do this the fibres were cut so that approximately 25~mm of fibre
protruded from the face of the connector.
A vacuum cup was then fitted to the connector and potting adhesive was
applied in the recess on the fibre-entry side of the connector.
The vacuum applied ensured that the adhesive travelled the length of
the fibre within the bore.
After the adhesive had been extruded through the bore, the vacuum was
removed and the recess `topped up' with additional adhesive. 
When the adhesive was in a stable state (i.e. had hardened) the station
was turned over and adhesive was applied to the front face of the
connector around the fibres to prevent any movement or vibration in
subsequent cutting and polishing operations.

\begin{figure}
  \begin{center}
    \includegraphics[width=0.45\textwidth]%
    {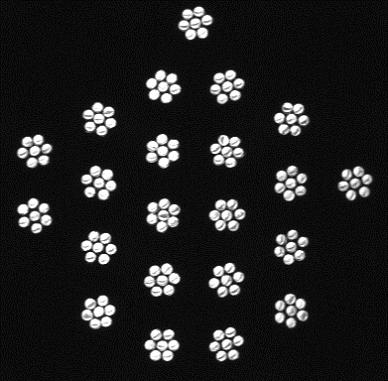}
  \end{center}
  \caption{
    Photograph of the face of one station connector.
    The doublet layer was illuminated with a defocused beam of
    light so that all fibres were excited.
    The bundles of seven fibres may clearly be seen.
  }
  \label{Fig:MechSciFiStat:OptiScan:Connect}
\end{figure}

Once the glue had cured, excess fibre and cured adhesive was cut at a
distance of 2--3~mm from the surface of the connector.
The remaining excess material was removed using a diamond-tipped
tool. 
The same tool was also used to skim the surface of the connector to
give the required high-quaility finish without the need for further
polishing steps which may have degraded the surface flatness.
Figure \ref{Fig:MechSciFiStat:OptiScan:Connect} shows a photograph of
the fibre bundles after polishing.

\subsubsection{Station quality assurance procedures}
\label{SubSubSect:StationQA}

\noindent{\bf Optical quality assurance procedures}

\noindent{Each bundle of seven scintillating fibres is expected to
produce a signal of approximately equal strength. 
The QA procedures applied during the station fabrication process
exploited this property to identify fibres that were damaged. 
}

\noindent{\it Fibre Doublet Scanning Machine}

\noindent{A weak 405~nm LED was used to excite the green fluorescent
emission of the 3HF dopant in each fibre. 
Each doublet layer was held on the vacuum chuck which was fitted to a
precision stage as described above.
The stage was used to step the line-focused LED source over the
doublet layer, exciting successive fibres in turn.
A sensitive video camera was used to view the end of each bundle of
seven fibres.
The response of the video camera was recorded and used to measure 
the transmitted light.
}

\begin{figure}
  \begin{center}
    \includegraphics[width=0.95\textwidth]%
    {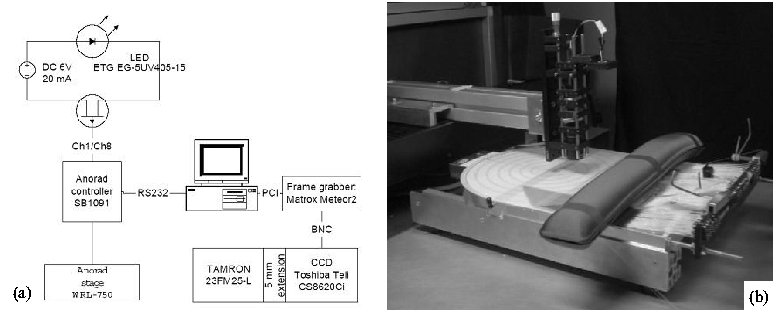}
  \end{center}
  \caption{
    (a) Schematic diagram of the QA system.
    (b) An overall view of the station assembly QA setup. 
    The optical system is attached to a linear stage and held over the
    plane of scintillating fibres to illuminate the channels in
    sequence. 
    The doublet layer is laid on top of a vacuum chuck, which is fixed
    in place by a guiding rail. 
    The bundled fibre ends are held up in order at the right hand side
    to be viewed by a CCD camera that is located outside this
    photograph. 
  }
  \label{Fig:MechSciFiStat:OptiScan:Schem}
\end{figure}

An Anorad WRL-750 linear-motor driven stage, with a total travel of
750 mm and a precision of 10 $\mu$m, carried an optical system to
illuminate the fibres.
The stage was driven from a PC via an RS232 link to an SB~1091
controller.
A CCD-based video camera, Toshiba CS8620Ci, was connected to
the PC via a Matrox Meteor II PCI card. 
The LED, type ETG-5UV405-15, was switched on and off under PC
control using one of the eight digital outputs of the SB~1091
controller. 
The scanning machine is shown schematically in figure
\ref{Fig:MechSciFiStat:OptiScan:Schem}a while figure
\ref{Fig:MechSciFiStat:OptiScan:Schem}b shows a photograph of the
machine as built.

\begin{figure}
  \begin{center}
    \includegraphics[height=0.95\textwidth]%
    {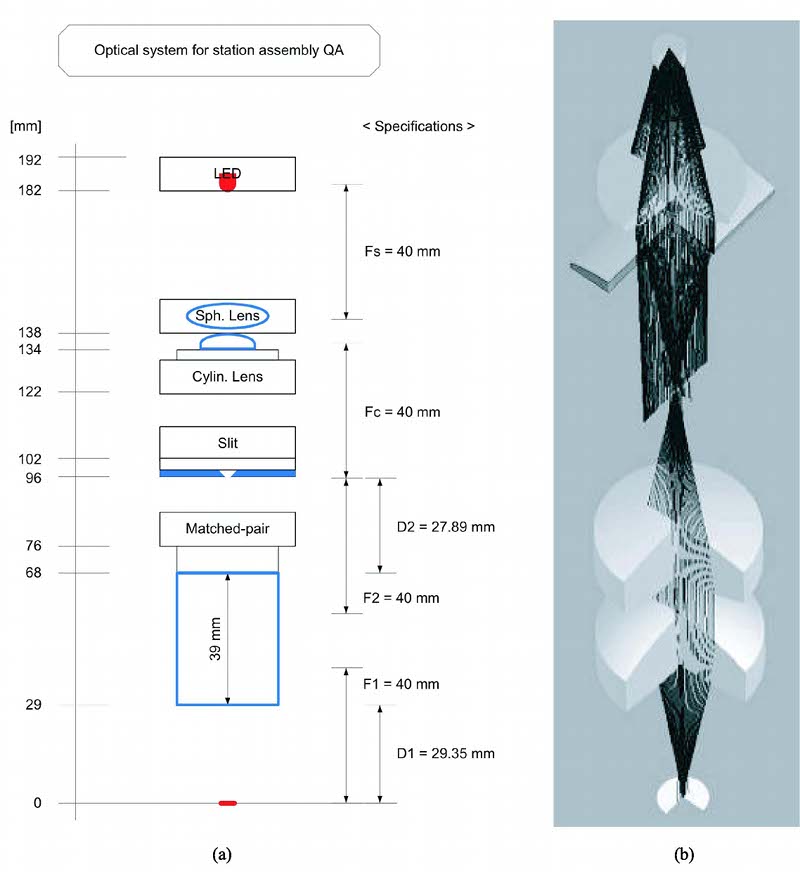}
  \end{center}
  \caption{
    (a) Optical layout of the fibre illuminator. 
    The total optical track is 184~mm from the LED to the image
    plane. 
    The LED is imaged onto a slit by a combination of a 40~mm focal
    length spherical lens and a 40~mm focal length cylindrical lens. 
    The slit (50 $\mu$m wide by 4~mm long) is re-imaged onto the 
    fibre by a matched achromatic doublet pair.
    (b) The diagram shows how the circular beam of light (represented
    here by ray fans from the LED at the top of the figure) is focused
    in only one plane by the cylindrical lens to produce
    a line focus cropped by the slit (tiny shaded bar at the centre),
    which is then projected on to the doublet layer (at the bottom) by
    the matched-pair lenses.
    This diagram was generated from a model of the optical system in
    the Zemax optical design package \cite{ZEMAX}. 
  }
  \label{Fig:MechSciFiStat:OptiScan:Optics}
\end{figure}

The LED produces a soft-edged, approximately circular, diverging beam
of light (illumination half-angle of $15^\circ$), which is projected
onto a narrow line by a two-stage optical system \cite{MICE_Note:OpticalSystem}.
The illumination system is shown schematically in figure
\ref{Fig:MechSciFiStat:OptiScan:Optics}a.
Figure \ref{Fig:MechSciFiStat:OptiScan:Optics}b shows a schematic of
the rays in the focusing system. 

The first stage of the optical system uses cylindrical optics to
transform the beam from the LED from a circular to a linear
cross-section.
As an LED is an extended source and includes features such as bond
wires, the projected line will have soft edges and internal structure
when in focus. 
A mechanical slit was therefore used to select a well-defined,
uniformly lit region close to the first focus and the image of this
slit was projected onto the doublet layer by the second stage of the
optical system which consisted of a matched pair of conventional
lenses.  
This approach also provided a clear working distance between the
optical system and the surface of the doublet layer, reducing the
risk of damage to the fibres during the scanning process.

Each doublet layer was scanned, after bundling was complete, to
ensure that the correct bundles of seven fibres had been grouped 
together.
This was achieved by performing a `counting-and-ordering check' in 
which the number of fibres illuminated was counted and the
illumination sequence was checked against that expected as the stage
passed across the doublet layer.
Any errors identified were corrected and the counting-and-ordering
check repeated.
The fitting of the station connectors only started after the correct
bundling had been achieved.

\begin{figure}
  \begin{center}
    \includegraphics[width=0.95\textwidth]%
    {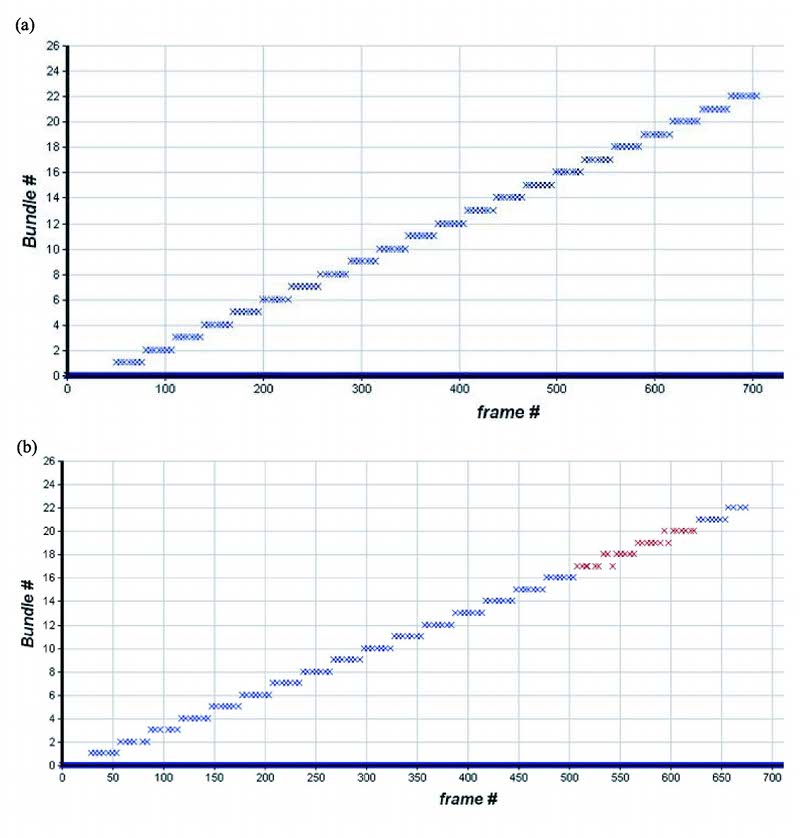}
  \end{center}
  \caption{
    Examples of the plots used to ensure that the fibres were threaded
    in the correct channel of the station optical connector.
    The bundle number is plotted against the frame number using the
    procedure described in the text.
    (a) Perfect sequencing of all the fibres in the connectors.
    (b) Single fibres swapped between bundles 17 and 18 and between
    bundles 19 and 20. 
  }
  \label{Fig:MechSciFiStat:OptiScan:Sequence}
\end{figure}

Once all the connectors had been fitted, the doublet layer was scanned
at a constant speed of $1.25$~mm.s$^{-1}$ with a video capture rate of
25~Hz to allow the fibre-sequence to be verified.
For each frame the $3 \times 3$ pixels around the centre of the illuminated 
fibre were summed. 
A search was then made through neigbouring frames to find the frame in
which the intensity of a particular fibre was largest and the frame
number for the intensity peak for each fibre was recorded. 
Plotting the frame number versus the fibre bundle number revealed
misplaced fibres.
Figure \ref{Fig:MechSciFiStat:OptiScan:Sequence}a shows an example of
the bundle/frame number correlation for perfect sequencing.
Figure \ref{Fig:MechSciFiStat:OptiScan:Sequence}b illustrates a fibre
sequence error.

\noindent{\bf Station acceptance procedure}

\begin{figure}
  \begin{center}
    \includegraphics[width=0.7\textwidth]%
      {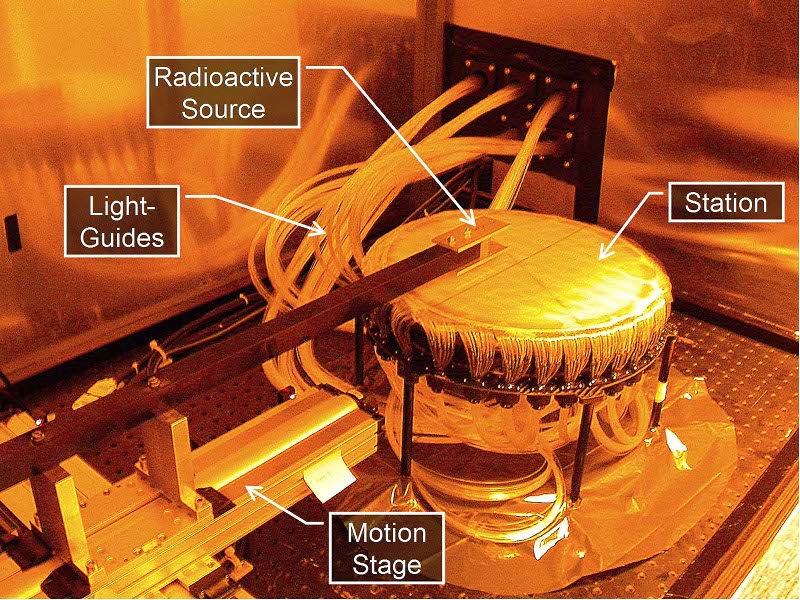}
  \end{center}
  \caption{
    A photograph of a tracker station mounted in the QA test stand. 
    The picture was taken from within the light-tight box which
    encloses the system.
    The internal light-guides can be seen connected at the station and
    the patch panel.
    The radioactive source was positioned in the holder which was
    mounted on the motion stage.
  }
  \label{Fig:StationAcceptanceProc:Overview}
\end{figure}

\noindent{Each completed station was required to undergo an additional
phase of QA testing in order to determine the uniformity of the light
yield across its active area.   
This was achieved by measuring the light output by inducing
scintillation in localised regions across the surface of the 
station using a point-like radioactive source. 
A photograph of the QA test stand in which this process was automated
is shown in figure \ref{Fig:StationAcceptanceProc:Overview}.
}

The station was mounted horizontally on a rigid support frame inside a
light-tight box. 
Scintillation light leaves the box via clear-fibre light-guides and is
converted to electrical signals via the `Visible Light Photon Counter'
(VLPC) system described in section \ref{Sect:Electronics}.
Digitisation of the analogue signals from the VLPCs was performed
using two prototype `Analogue Front End' (AFE) boards developed by the
D\O{} collaboration.
A collimated radioactive source was held at a fixed distance of
approximately 1~mm from the surface of the station. 
The source could be moved to any location in the plane of the active
area by a programmable 2D motion stage, with an absolute position
accurate to better than $\pm$5~$\mu$m. 

Initially, a $^{90}$Sr beta source was selected for the QA test stand. 
It was expected that individual beta particles which passed through
all three planes of a station would provide a hit in each of the three
doublet layers, providing a clean signature (referred to as a
`triplet') from which the position of the source could be
reconstructed. 
The QA analysis would then compare the reconstructed position with the
position set by the 2D stage. 

\begin{figure}
  \begin{center}
    \mbox{
      \subfigure[]{
        \includegraphics[width=0.45\textwidth]%
          {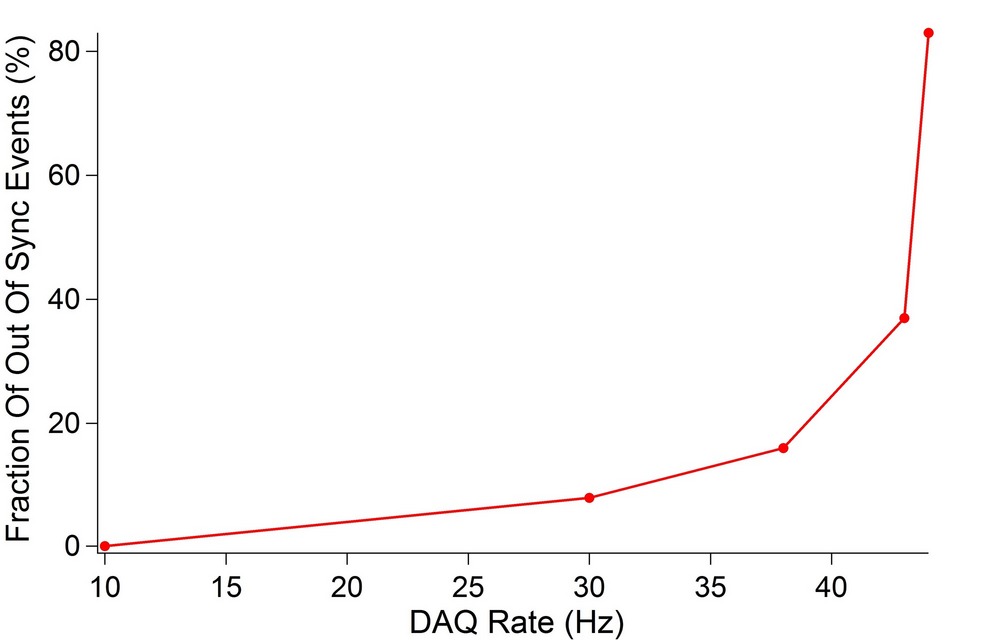}
      }
      \subfigure[]{
        \includegraphics[width=0.45\textwidth]%
          {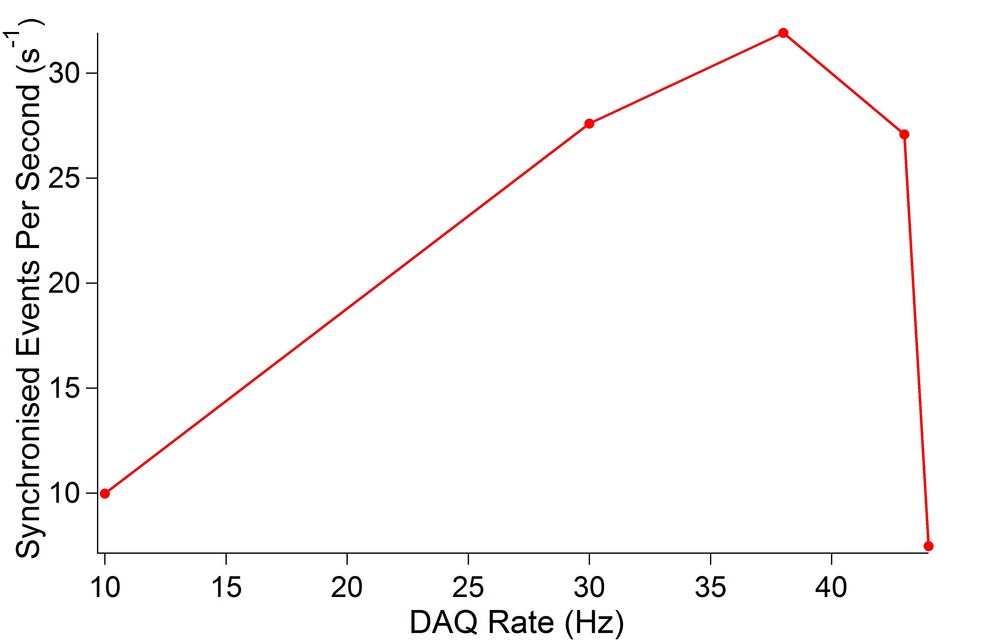}
      }
    }
  \end{center}
  \caption{
    (a) A graph showing, as a function of DAQ trigger rate, the
    fraction of events acquired from the QA DAQ system in which the
    data from the two AFE boards are not correctly synchronised.
    (b) A graph of the synchronised event rate versus DAQ event rate.
  }
  \label{Fig:StationAcceptanceProc:SyncEvents}
\end{figure}

It was found that the overall rate at which
data could be read out from the two AFE boards in the DAQ system
was limited by hardware synchronisation issues. 
Due to time constraints and limited availability of equipment, it was
necessary to run the AFE boards in `self triggered' mode rather than
with a common external trigger (as will be the case during operation
in MICE).
In this mode, each board independently determines whether an `event'
should be triggered by comparing the `OR' of all the signals with a
predefined threshold.
At high data rates, the AFE boards may each trigger on scintillation
light produced by different successive physical stimuli and thus lose
synchronisation, leading to invalid data. 
This phenomenon was investigated by varying the internal AFE
thresholds to modify the number of signals accepted as triggers.
The fraction of unsynchronised `events' increased exponentially with
DAQ trigger rate, as shown in figure
\ref{Fig:StationAcceptanceProc:SyncEvents}(a), limiting the maximum
effective readout rate to $\sim$38 Hz (see figure
\ref{Fig:StationAcceptanceProc:SyncEvents}(b)).

In addition to the hardware-imposed rate limits, it was discovered
that the mean energy of the $^{90}$Sr beta particles was below that
required for penetration through all station planes. 
Due to absorption and scattering, the number of detected particle hits
in each plane decreased with distance from the source such that only
$\sim$4\% of recorded events corresponded to a triplet. 
This resulted in a data rate insufficient for the routine
characterisation of tracker stations using only triplet events.
It therefore became necessary to consider all events when calculating
light yield. 
The triplet data acquired during the preliminary testing phase were
used to calibrate the alignment of the stations with the motion stage
coordinate system such that all data values could be mapped to a
specific source position using the associated channel number, rather
than by reconstructing a complete particle trajectory. 
Since the triplet rate produced by the beta source was not
appropriate for QA purposes, the $^{90}$Sr was replaced by a $^{57}$Co
gamma source in the final system.
From initial $^{57}$Co measurements it was determined that 50k events
were required at each source location to determine the light yield
with sufficient precision.

\begin{figure}
  \begin{center}
    \includegraphics[width=0.60\textwidth]%
      {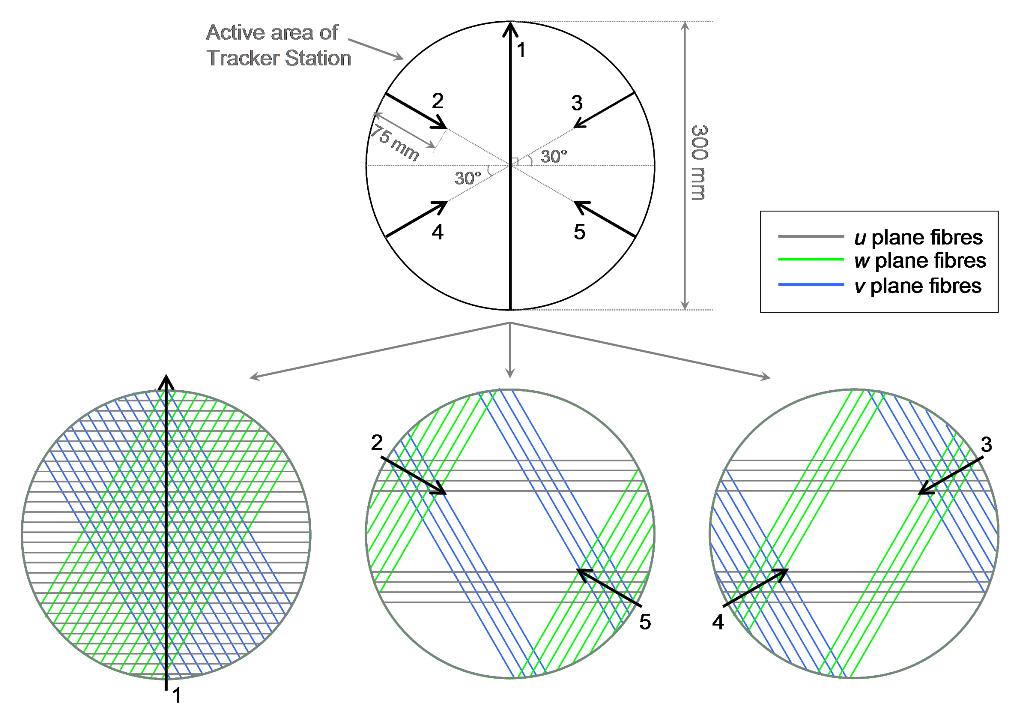}
  \end{center}
  \caption{
    A diagram showing the path of the source as it is scanned across
    the active area of a tracker station during the QA test
    procedure. 
    Scan 1 is perpendicular to the fibres in plane \textit{u}, and
    extends across the entire active region. 
    Scans 2 and 5 are perpendicular to the fibres in plane \textit{w},
    while scans 3 and 4 are perpendicular to those in plane
    \textit{v}; each of these scans has a nominal length of one
    quarter of the active area diameter. 
    The lower portion of the figure illustrates the manner in which
    subsets of the \textit{u}, \textit{v} and \textit{w} plane fibres
    were illuminated by the source during each scan (not drawn to
    scale).
  }
  \label{Fig:StationAcceptanceProc:ScanPattern}
\end{figure}

In defining the path that should be traced by the source during a QA
test, it was necessary to compromise between the number of channels
to be characterised and the overall duration of the scan. 
Ideally, light yield would be measured with the source placed at a
number of points along the length of every fibre channel in each
station plane, but the low data rate and large sample size required at
each source position rendered this approach impractical.
The most efficient scan pattern was therefore defined as the shortest
possible path which crosses each channel at least once. 
The presence of the mirror on one end of the fibre implies that a
fault or break in a fibre will reduce the overall light yield
regardless of the position at which scintillation occurred.
Hence, one measurement per channel was judged to be sufficient to
identify abnormal fibres for which further investigation was required. 
The format of the QA scan pattern is shown in figure
\ref{Fig:StationAcceptanceProc:ScanPattern}. 

Despite simplifying the scan path, the limited time available for the
completion of the initial phase of QA testing prior to the assembly of
the first tracker was such that only a subset of the channels of each
station could be routinely evaluated. 
It was ultimately necessary to test each station within a period of
one day, nominally corresponding to two consecutive 8-hour shifts by
the test operatives. 
Excluding the time required to prepare and install a station in the QA
stand (and to remove it following a test run) $\sim$11 hours were
allocated for data acquisition. 
With a requirement of 50k events per source position and a maximum
data rate of $\sim$38~Hz, this enabled measurements to be taken at 28
discrete locations during the scan, at intervals of 27.4~mm 
($\sim$18 channels). 
However, had any evidence of light yield non-uniformity been detected
at this stage, higher resolution studies of individual channels in the
affected area would have been made. 
No evidence of such a non-uniformity was observed.

%
\subsection{Light-guides and optical connectors}
\label{SubSect:LightGuidesOpticalConnectors}

\subsubsection{Light-guides}
\label{SubSubSect:LightGuides}

Clear-fibre light-guides are required to transport light from the
scintillating-fibre stations to the VLPC readout system.
Kuraray clear polystyrene, round, s-type, 1.05~mm fibre was
chosen for the light-guides \cite{Kuraray}. The attenuation length 
of the fibre was measured to be 7.6~m and thus, to reduce the light loss
in propagation as far as possible, the length of the clear-fibre run
from station to VLPC cassette was specified to be 4~m.

The light-guide, a bundle of 128 clear fibres, is composed of an
internal section, which connects between the station and the patch
panel, and an external section, which connects between the patch panel
and the VLPC cassette.
The external light-guide is equipped with a flexible tube which
shields the fibres from ambient light and also protects the fibres
from any damage during handling and assembly.  
The length of each light-guide is summarised in table
\ref{light_guide_length}. 
Optical grease was applied to all optical connectors when mating the
light-guides in order to minimise light yield loss due to Fresnel
reflection.
\begin{table}
\caption{
    Light-guide lengths. 
    The serial number is a tracking number marked on the connectors. 
    Tracker 1 is for the upstream spectrometer and tracker 2 for the
    downstream spectrometer. 
    Station 5 is located nearest to the patch panel, and station 1 is
    furthest from the patch panel.
  }
  \begin{center}
    \begin{tabular}{ccccc}
      Serial Number & Tracker & Station & Internal & External \\
      \hline
      6-10 & 1 & 5 & 1500 mm & 2500 mm \\
      16-20 & 1 & 4 & 1950 mm & 2050 mm \\
      21-25 & 1 & 3 & 2250 mm & 1750 mm \\
      26-30 & 1 & 1 & 2600 mm & 1400 mm \\
      31-35 & 1 & 2 & 2400 mm & 1600 mm \\
      36-40 & 2 & 5 & 1500 mm & 2500 mm \\
      41-45 & 2 & 4 & 1950 mm & 2050 mm \\
      46-50 & 2 & 3 & 2250 mm & 1750 mm \\
      51-55 & 2 & 2 & 2400 mm & 1600 mm \\
      56-60 & 2 & 1 & 2600 mm & 1400 mm \\
      \hline
    \end{tabular}
  \end{center}
  
  \label{light_guide_length}
\end{table}

\begin{figure}
  \begin{center}
    \includegraphics[width=0.75\textwidth]%
    {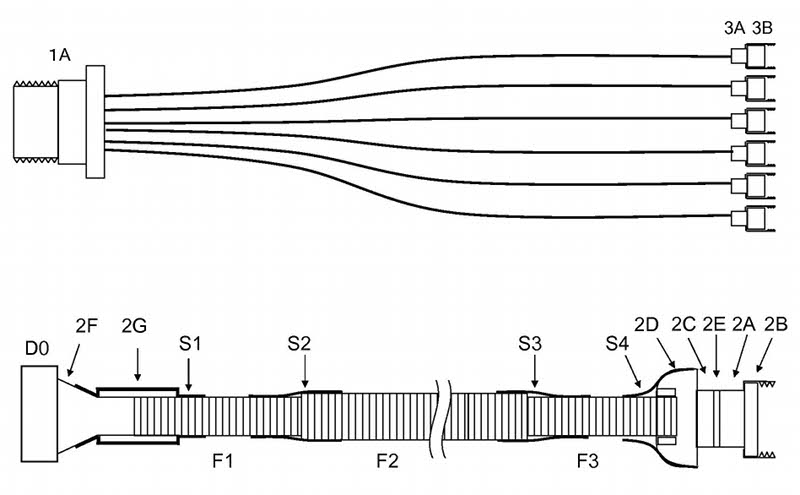}
  \end{center}
  \caption{
    Drawings for internal light-guide (top), and  external light-guide
    (bottom). 
    1A, 3A, D\O{}, 2A are optical connectors; 3B, 2B are locking
    rings.
  }
  \label{light_guide}
\end{figure}

\begin{figure}
  \begin{center}
  \includegraphics[width=0.75\textwidth]%
    {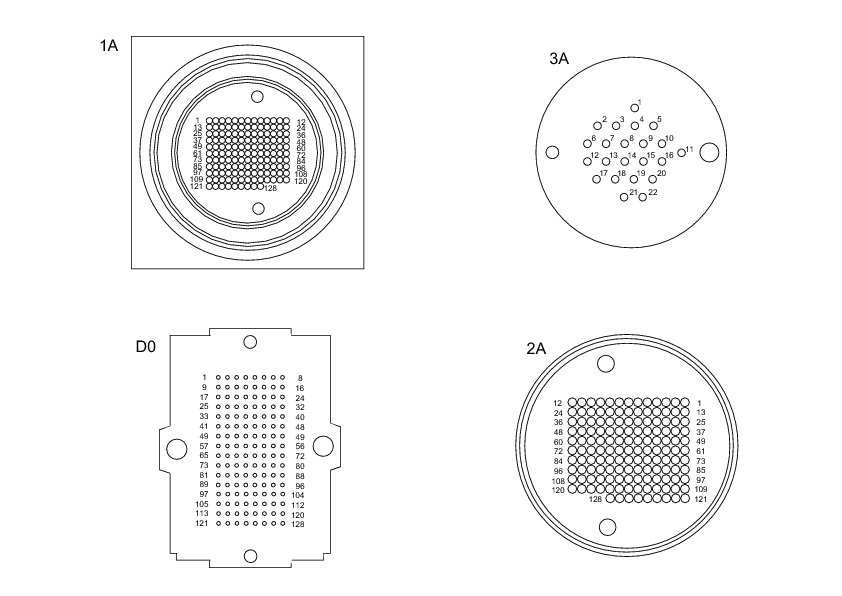}
  \end{center}
  \caption{
    Schematic view of optical connectors. 
    The connector 1A (top-left) and 3A (top-right) are for the internal
    light-guide, D\O{} (bottom-left) and 2A (bottom-right) for the external
    light-guide.
  } 
  \label{optical_connector}
\end{figure}

Each internal light-guide (figure \ref{light_guide} top) has one
128-fibre `patch-panel' connector (1A) and six 22-fibre station
connectors (3A).
The connector 1A (figure \ref{optical_connector} top-left) has 128
holes which match with those of the connector 2A (figure
\ref{optical_connector} bottom-right) of the external light-guide.
The connector 3A (figure \ref{optical_connector} top-right) has 22
holes configured as shown.  
These connect to the station connector and are locked by the collar
3B.

The external light-guides (figure \ref{light_guide} bottom) use a
128-fibre optical connector, developed by D\O{}, at the end that
interfaces to the VLPC cassette and a connector assembly (2A-E) which
mates to connector 1A of the internal light-guide. 
Flexible tubes F1, F2, and F3, and heat-shrink tubes S1, S2, S3, and
S4 form the body casing of the external light-guide.
A light-tight injection-moulded boot (2F) is glued into the rear of
the D\O{} connector and flexible tube F1 is attached to the boot using
sleeve 2G with adhesive and heat-shrink tubing S1. 
This makes the assembly fully light tight and adds strength. 
The flexible tubes F1, F2, and F3 are fixed with heat-shrink tubes S2
and S3.
The locking ring 2B is used for securing the connector 2A to the
connector 1A of internal light-guide at the patch panel. 
The optical connector D\O{} (figure \ref{optical_connector} bottom-left)
has 128 holes, in a $16 \times 8$ arrangement which make contact
with the connector at the VLPC cassette. 
The optical connector 2A (figure \ref{optical_connector} bottom-right)
has 128 holes in an arrangement as shown, which mates with connector
1A of the internal light-guide. 

\subsubsection{Light-guide fabrication}
\label{SubSubSect:LightGuidesFabrication}

Five-metre lengths of fibre were cut from the spool on which they were
delivered by the manufacturer.
A bundle of fibres was then connected to cookies at each end in order
to perform the first transmittance measurement (see below).
Fibres passing the transmittance test were cut into two pieces, one for
the internal light-guide and the other for the external one.
Next, optical connectors were attached at each end of the bundle and
glued in position to allow the first polishing step to be performed. 
A second transmittance measurement was then performed.
The connectors that passed this test were polished using a diamond
cutter.

\subsubsection{Light-guide quality assurance}
\label{SubSubSect:LightGuidesQA}

\begin{figure}
  \begin{center}
    \includegraphics[width=0.75\textwidth]%
    {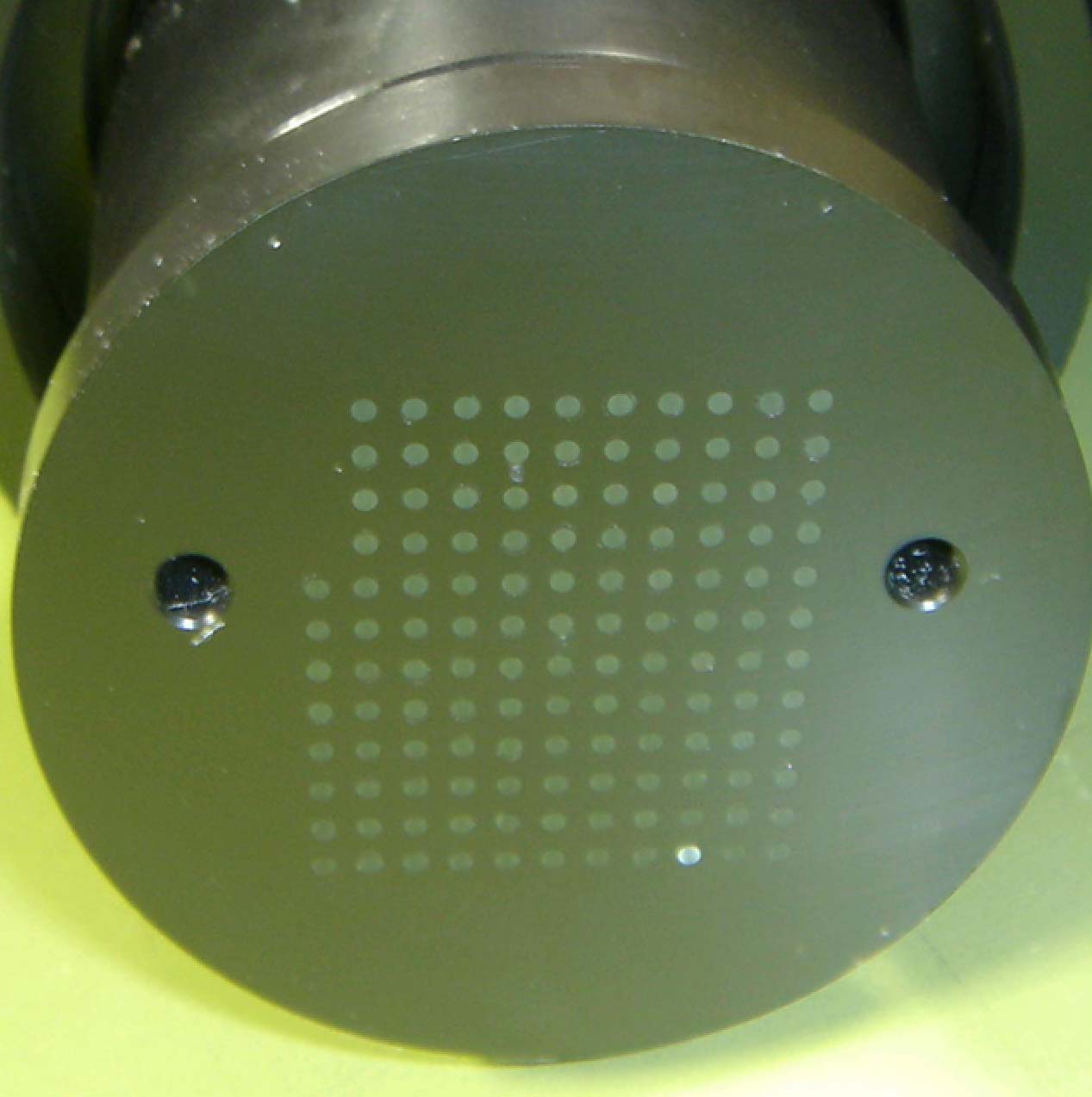}
  \end{center}
  \caption{
    Photograph of the connector surface taken with strobe illumination. 
    Reflected light is seen in a fibre at the bottom.
  }
  \label{reflection}
\end{figure}

\begin{figure}
  \begin{center}
    \includegraphics[width=0.75\textwidth]%
    {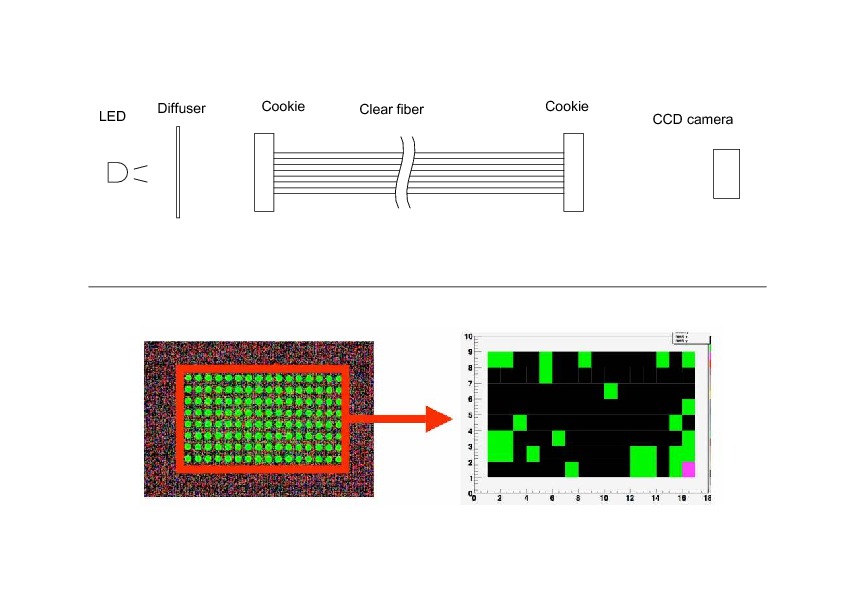}
  \end{center}
  \caption{
    Setup for transmittance measurement (top). 
    The light from the green LED passes through a diffuser and
    illuminates the fibres.
    The intensity of the transmitted light is measured using a CCD
    camera.
    The CCD image of transmitted light (bottom-left) is analysed and 
    the intensity of the green component in each fibre is checked.
    The intensity of each channel, relative to the average for the
    light-guide under test, is plotted channel by channel (bottom
    right).
    The red pixel indicates a channel with a transmitted light
    intensity 8\% lower than the average.
}
  \label{qa_setup}
\end{figure}

After a brief inspection of the fibre connections, three quality
assurance steps were carried out.
The first step was to search for evidence of kinks or cracks in the
fibres produced in handling the light-guides during manufacture. 
The surface of each connector was exposed to a strobe light and the
reflected light was captured in a photograph taken with a digital
camera.
An example of such a photograph, showing evidence of damage in one
fibre, is shown in figure \ref{reflection}.
The damaged channel is clearly visible in the photograph where the
injected light is reflected from the damage point along the fibre.
The second QA step was to measure transmittance in the fibres. 
The bundle of fibres was illuminated using a diffuse source of green
light (see figure \ref{qa_setup}).
The light transmitted along the light-guide was observed with a CCD
camera.
The variation in light transmission from fibre to fibre was found to
be less than 10$\%$. 
The fraction of fibres with transmission less than 90$\%$ (of the
average) was less than 0.7$\%$. 
Finally, the fibre mapping at each connector was checked.

\subsubsection{Optical connectors}
\label{SubSubSect:OpticalConnectors}

\begin{figure}
  \begin{center}
    \includegraphics[width=0.95\textwidth]%
    {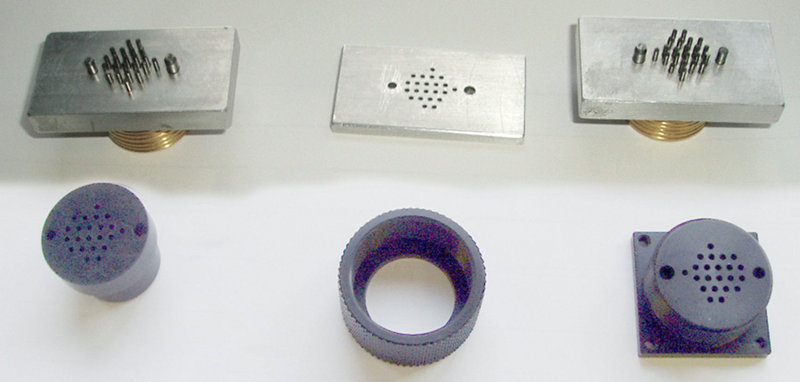}
  \end{center}
  \caption{
    The station connector gauge set. 
    The gauges are arranged in a row at the top of the photograph.
    The left- and right-hand gauges were used to check the connector
    part shown directly below.
    The `gauge gauge' shown in the centre of the top row is used to
    ensure that the pins in the two gauges are precisely matched.
  }
  \label{Fig:StnConnGuageSet}
\end{figure}

The station optical connector is shown in figure
\ref{Fig:StnConnGuageSet} together with the various gauges used in the
station-connector QA.
The connector has 22 1.05~mm-diameter holes, drilled to take the
fibres, and was machined from black Delrin. 
The alignment dowels have different diameters, 2~mm and 2.5~mm, to
prevent any possibility of mating the connector pair in the wrong
orientation.  
To ensure that the two halves of the connectors have no misalignment
(which would cause light loss) a set of gauges was manufactured.  
These consisted of a plate in which a series of dowels were
accurately positioned at the locations where the holes in the
connector should be.
A separate gauge was made for each half of the connector.
To check the accuracy of the two gauges, a precisely drilled plate was
produced.
All of the connectors were checked using one of these gauges, and any that
failed were discarded.

\begin{figure}
  \begin{center}
    \includegraphics[width=0.95\textwidth]%
    {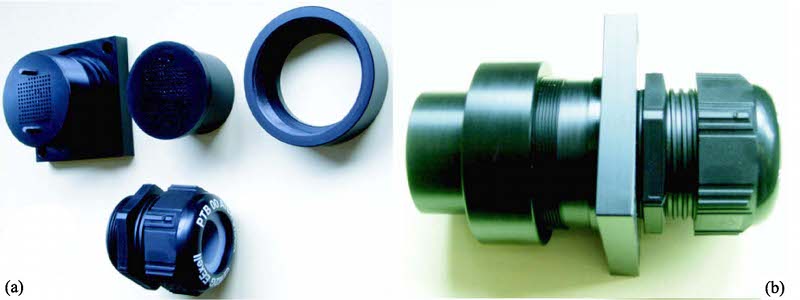}
  \end{center}
  \caption{
    Components of the patch-panel connector.
    (a) The components before assembly.
    The bulkhead connector carries a square flange to carry an O-ring
    and to allow the connector to be held with screws onto the patch 
    panel.
    The mating connector with locking ring is shown as well as the
    strain relieving piece (the lightly shaded ring) that is attached to the external
    light-guide.
    (b) The assembled patch-panel connector.
  }
  \label{Fig:PtchPnlConn}
\end{figure}

Figure \ref{Fig:PtchPnlConn} shows the patch-panel connector
components.
The connector bearing the square flange is fitted to the internal
light-guide and incorporates an O-ring to make a gas seal.
The patch-panel connector is machined from black Delrin. 
There are six station connectors to each patch-panel connector, though
not all of the channels in all the station connectors are used.  

\begin{figure}
  \begin{center}
    \includegraphics[width=0.45\textwidth]%
    {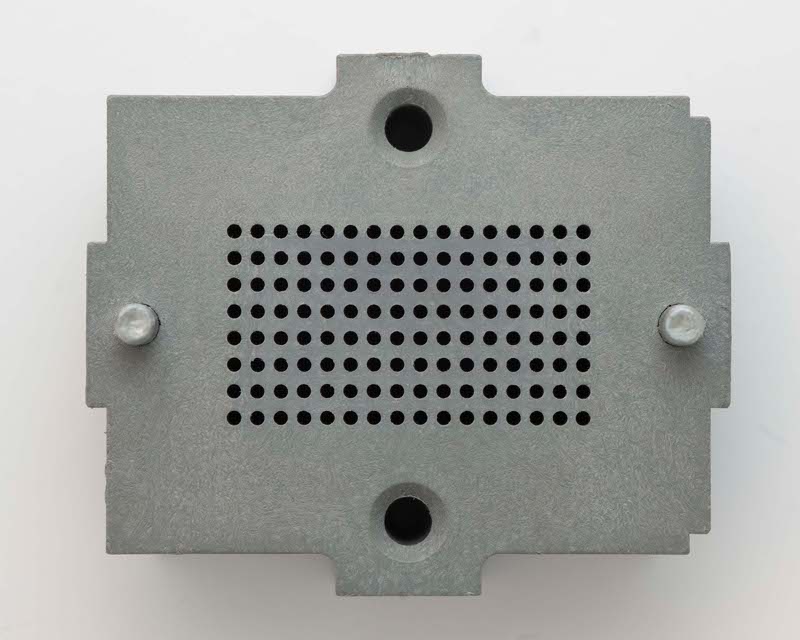}
  \end{center}
  \caption{
    D\O{} warm-end optical connector.
  }
  \label{Fig:D0NConn}
\end{figure}

The optical signals are fed from the patch-panel to the VLPC system via
the external light-guides, each containing 128 fibres. 
The light-guides terminate in the D\O{} warm-end connector shown in
figure \ref{Fig:D0NConn}.
This is an injection-moulded part, also made from Delrin. 
The figure shows the 128 holes for fibres, two holes (left/right) for 
alignment pins and two holes (top/bottom) for threaded inserts. 
Light-guide fibres of 1.05 mm diameter are used while the D\O{} VLPC cassettes
use fibres of 0.965 mm diameter. 
This mismatch results in a light loss of $\sim$15\%.

%
\subsection{Tracker assembly and integration}
\label{SubSect:TrackerAssembly}

\subsubsection{Assembly}

The stations to be used in the final trackers were chosen on the basis
of the results obtained in the QA exposure of the stations to the
radioactive source.
The stations that make up a tracker are held in position using four
carbon-fibre space frames.
The relative positions of the stations were determined using the
following assembly procedure.

\begin{figure}
  \begin{center}
    \includegraphics[width=0.70\textwidth]%
    {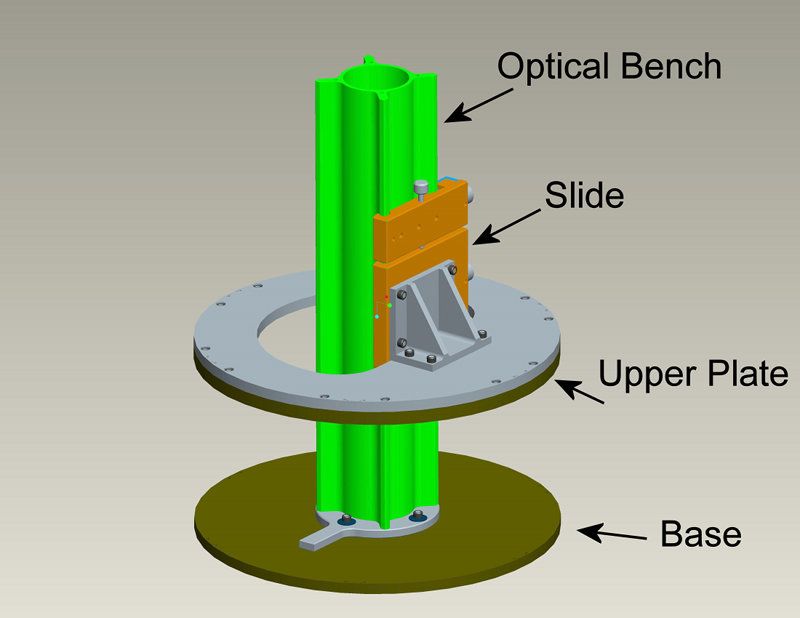}
  \end{center}
  \caption{
    Engineering model of the gluing jig showing the base plate
    (labelled `base'), the optical bench and slide, and the upper
    plate. 
    The space frame was held in position between the upper plate and
    the base plate while the glue was allowed to cure.
  }
  \label{Fig:TrkrAssmbly:GluingJig}
\end{figure}

Each space frame was assembled in a gluing jig (see figure
\ref{Fig:TrkrAssmbly:GluingJig}) which consisted of an
optical bench fixed to a base that included dowel holes matching those
in the station bodies. 
A similar plate was mounted on an optical slide and the distance between
this plate and the base plate could be adjusted and set precisely.
The distance between the two plates was checked, not only to ensure
the correct inter-station distance, but also to ensure that the two
plates were parallel.

The four sections of the carbon-fibre space frame consist of 25
components: 12 feet; 6 inter-station tubes; 6 `W' sections; and a
stiffening tube.  
Each section was assembled between the two plates of the gluing jig
(figure \ref{Fig:TrkrAssmbly:GluingJig}) and bonded.    
When the assembly had fully cured it was removed from the jig and
measured; the jig was then reset to the next inter-station pitch.

\begin{figure}
  \begin{center}
    \includegraphics[width=0.60\textwidth]%
    {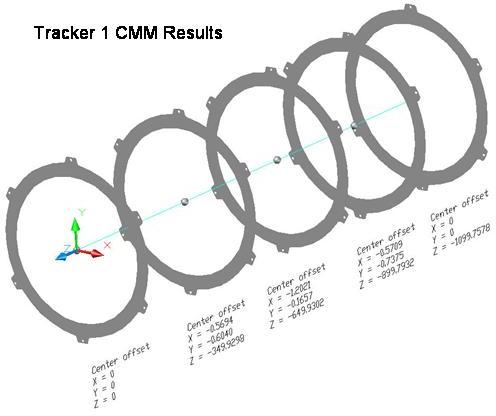}
  \end{center}
  \caption{
    Representation of the measured positions (in mm) of the five stations of
    the first MICE tracker.
    The centres of stations 2, 3, and 4 were determined relative to
    the centres of stations 1 and 5 and the $z$ position of stations
    2, 3, 4, and 5 were determed relative to the position of station 1
    as described in the text.
  }
  \label{Fig:TrkrAssmbly:CMMMeasurements}
\end{figure}

It was a simple matter to assemble the stations and space frames into
a complete tracker as all the joints are doweled.  
When a complete assembly was finished it was measured using a CMM to
verify the position of each station relative to the axis of the
tracker. 
The axis of the tracker was taken to be the line joining the centre of
the active area of station 1 with the centre of the active area of
station 5. 
The centres of stations 2, 3 and 4 were then determined and reported
as deviations from this axis.
The distance between the stations was determined by taking the mean of
four position measurements taken from the machined face of the station
flange. 
Figure \ref{Fig:TrkrAssmbly:CMMMeasurements} shows the position 
of each station in the first MICE tracker.
The precise measurements of the station positions have been used in the
reconstruction of the data from the first MICE tracker that is
presented below.

\subsubsection{Integration}

\begin{figure}
  \begin{center}
    \includegraphics[width=0.60\textwidth]%
    {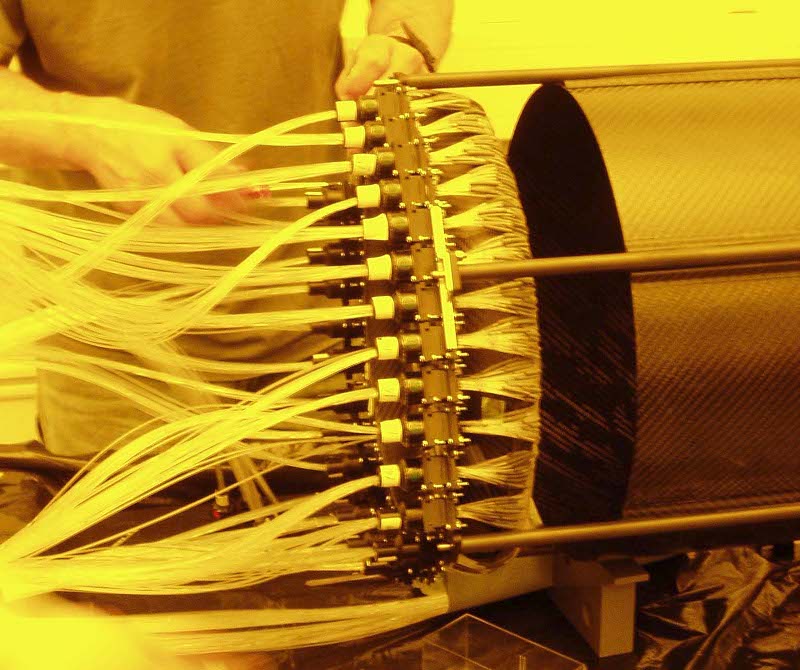}
  \end{center}
  \caption{
    Photograph of assembled tracker showing the internal light guides
    being attached to the support frame by which they are supported
    during the installation of the tracker into the bore of the
    solenoid. 
    In the picture, the tracker is being prepared for installation in
    the light-tight carbon-fibre tube used to house the tracker in the
    cosmic ray test stand.
  }
  \label{Fig:TrkrAssmbly:WgAttchmnt}
\end{figure}

The installation of the tracker into the spectrometer solenoid bore
will have to be carried out in UV-filtered light to avoid any
degradation of the scintillating fibres.
To achieve this, a blackout tent will be erected around the end of the
solenoid and made light tight. 
Specially filtered lighting will then be installed in the tent
allowing the tracker to be removed from its light-tight container and
placed on the installation platform.   
The platform is adjustable to enable the tracker to be aligned with
the bore of the solenoid.  
When on the platform, the frame that will support the internal
light-guides during the installation will be attached to the 
tracker via a flexible coupling to ensure that no stresses are
transferred to the tracker during the fitting of the internal 
light-guides (see figure \ref{Fig:TrkrAssmbly:WgAttchmnt}). 
The light-guide connections will be checked to ensure that the
light-guides are correctly positioned and labelled.

\begin{figure}
  \begin{center}
    \includegraphics[width=0.60\textwidth]%
    {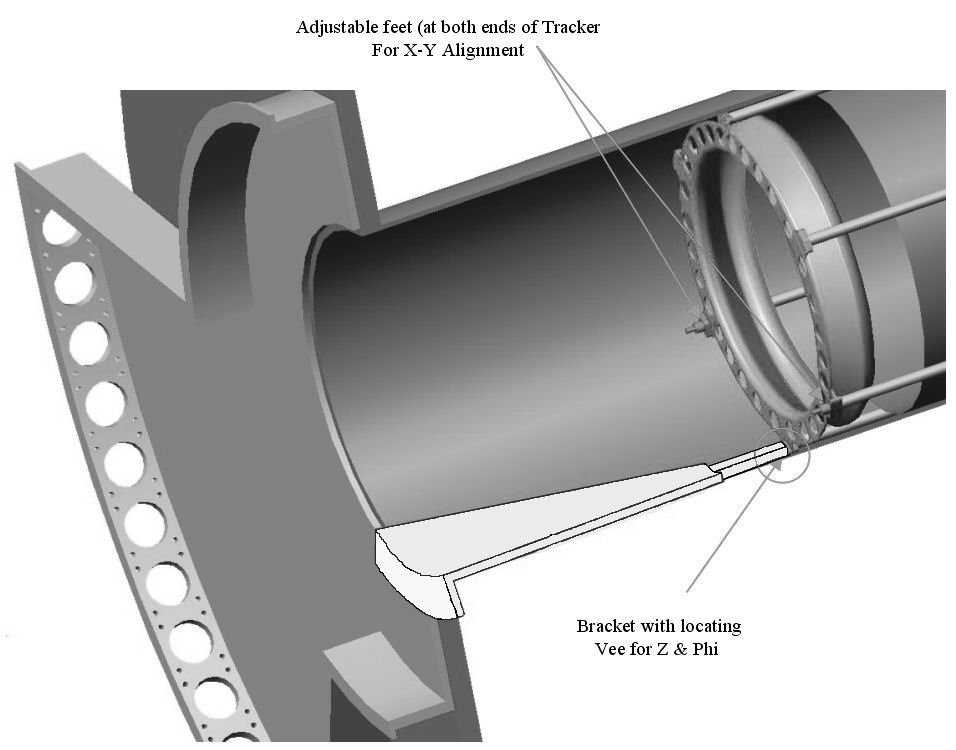}
  \end{center}
  \caption{
    Tracker alignment mechanism.
    The lighter part shown will be dowelled to the spectrometer
    solenoid end plate.
    At the narrow end of the part, inside the bore of the solenoid, a
    locking mechanism will be used to locate the tracker in $z$ and in
    azimuth.
  }
  \label{Fig:TrkrAssmbly:StnAlignMech}
\end{figure}

With the internal light-guides attached, the tracker assembly
(consisting of the tracker, the light-guide support, and the
light-guides) is ready for installation.
If necessary, the tracker assembly can be stored in an extended
light-tight storage tube for installation at a later date. 
For installation, the complete tracker assembly is slid into the
solenoid bore to the correct `$z$' position.
The tracker will sit on four adjustable feet, two at each end.
The adjustable feet will be used to align the tracker with the
magnetic axis of the solenoid. 
Once this has been done, the location bracket will be fitted.
The location bracket locks the tracker in its $z$ and azimuthal
positions (see figure \ref{Fig:TrkrAssmbly:StnAlignMech}).

\begin{figure}
  \begin{center}
    \includegraphics[width=0.75\textwidth]%
    {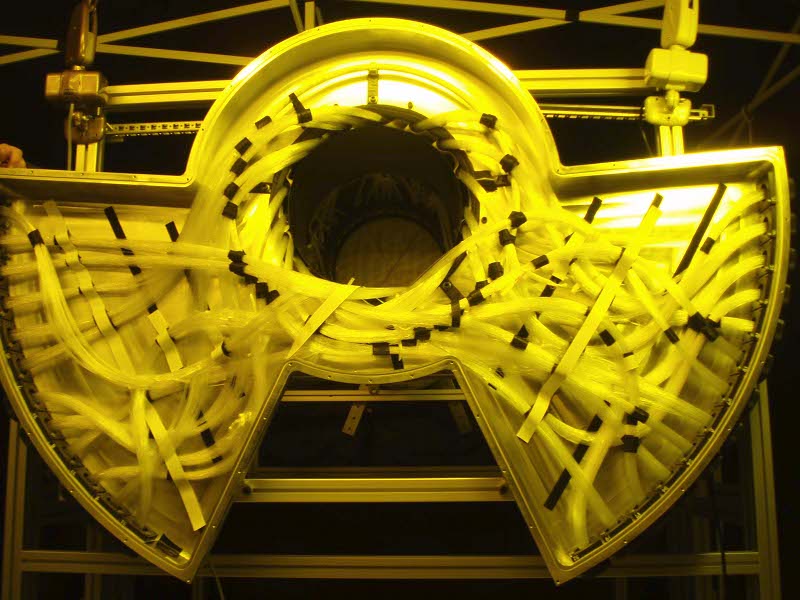}
  \end{center}
  \caption{
    Photograph of the internal light-guides supported within the patch
    panel.
    The photograph was taken at the end of the assembly of the first
    tracker in the cosmic test stand.
    The light-guides are held out of the tracking volume using a
    sprung carbon-fibre collar inserted into the the bore.
  }
  \label{Fig:TrkrAssmbly:IntlWGLayout}
\end{figure}

The light-guides will now be removed one-by-one from the support frame
and fitted into the patch panel.
This is a procedure that requires great care to ensure that the fibres
are not damaged and a collar will be fitted to ensure that the bore is
kept clear (see figure \ref{Fig:TrkrAssmbly:IntlWGLayout}).
Finally, the patch-panel cover will be fitted and, if the external
light-guides are to be fitted at a later stage, the patch panel 
connectors will have their light-tight caps fitted. 

%
\section{Photon Detection System}
\label{Sect:PhotonDetection}

The photo-detection system for the MICE trackers is built on the
success of the system built for the D\O{} experiment\cite{D0NIM}.
The D\O{} Central Fiber Tracker (CFT) is read out using Visible Light
Photon Counters (VLPCs) \cite{VLPC,VLPC1}. 
The extremely high quantum efficiency of the VLPCs was essential for
the performance requirements of the D\O{} experiment to be met and, for
MICE, has allowed a very aggressive fibre-tracker design based on
350~$\mu$m diameter fibres to be developed.
The MICE trackers will use D\O{} CFT readout and front-end electronics
(on loan from the D\O{} experiment).

\subsection{Visible Light Photon Counter - VLPC}

\begin{figure}
  \begin{center}
    \includegraphics[width=0.45\textwidth]%
    {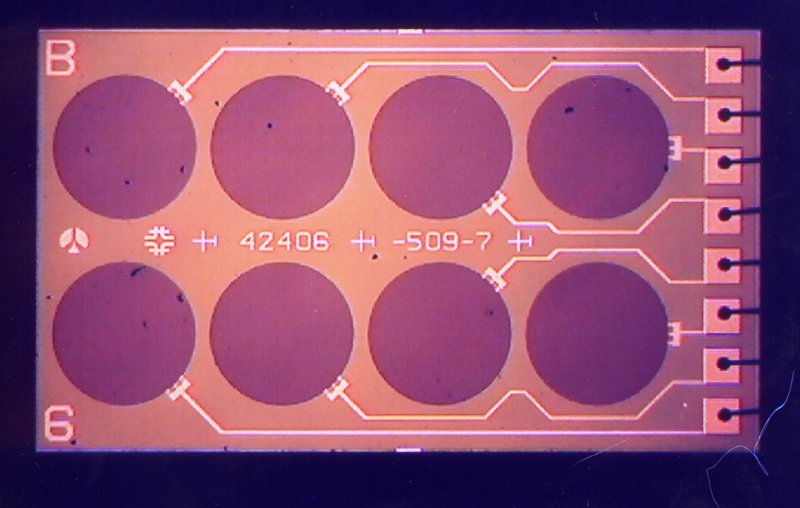}
  \end{center}
  \caption{Photograph of the 8-element VLPC array.}
  \label{VLPC600a}
\end{figure}

The light produced from charged particles passing through the
scintillating fibres is piped along the fibres and thence, via the
clear-fibre light-guides, to the VLPCs, which are housed in cassettes
(see below) and which convert the light into an electrical signal.
The VLPC is a silicon-avalanche device operated at cryogenic
temperature. 
It is a development of the Solid State Photomultiplier, an
impurity-band silicon avalanche photo-detector.  
It has undergone six design iterations, specified as HISTE~I to 
HISTE~VI.  
HISTE~VI, the version used in the D\O{} CFT, and thus the version that
is used in MICE, is an eight element array in a 2-by-4 geometry,
(figure \ref{VLPC600a}) with each pixel in the array having a diameter of 1~mm. 
The HISTE VI operational parameters are: quantum yield, $>$ 0.75; 
gain ($G$), 20,000--60,000; operating temperature, 9~K; and operating
bias, 6--8~V. 

VLPCs are fabricated using an epitaxial growth process which produces
a series of doped and undoped silicon layers on highly-doped silicon
substrate wafers. 
Each eight-pixel chip (figure \ref{VLPC600a}) is soldered to an
aluminium nitride substrate.
The outputs from individual pixels are wire-bonded to individual
contact pads on the substrate.
VLPCs from different batches exhibited a large variation in gain
due to non-uniformities in the production process \cite{bross}.
Due to these variations, the VLPCs were selected from individual
wafers and sorted into like-gain batches with small gain dispersion.
MICE will be using cassettes with gains of approximately 40k and
approximately 20k. 

\subsection{VLPC cassettes and cryostats}

\begin{figure}
  \begin{center}
    \includegraphics[width=0.95\textwidth]%
    {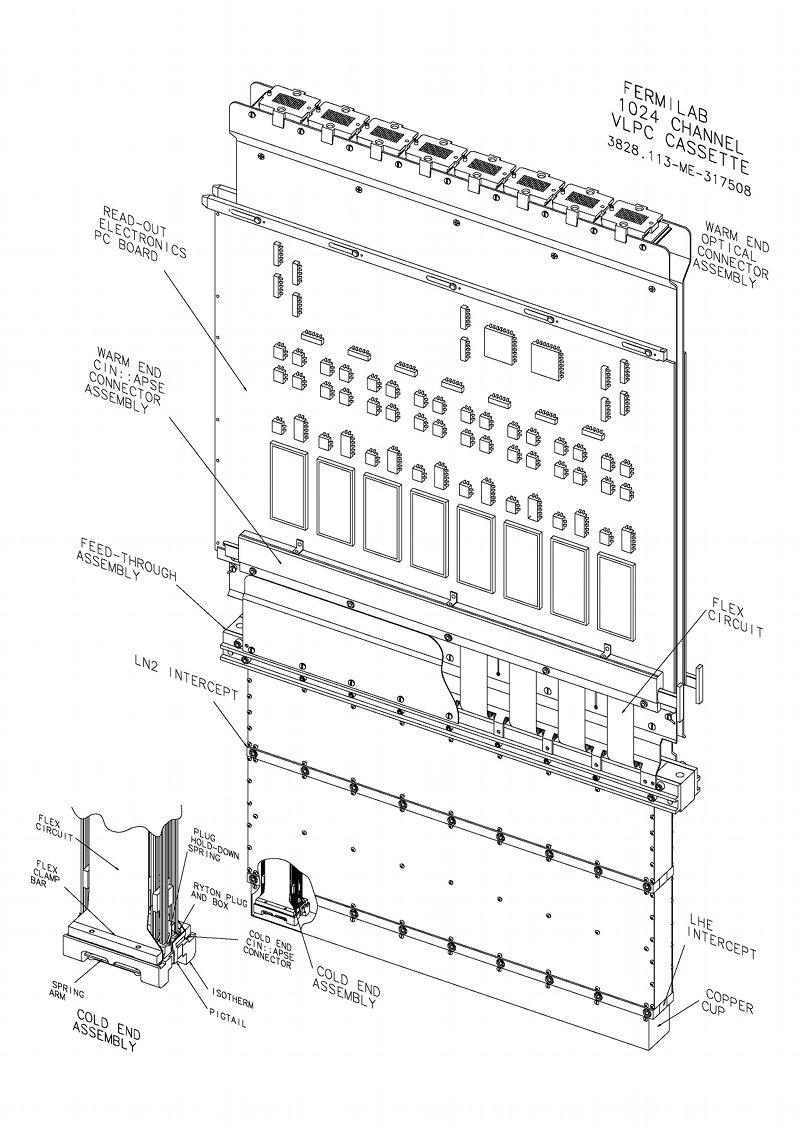}
  \end{center}
  \caption{
    A VLPC cassette supporting AFE readout boards as viewed from the
    left side. 
    The VLPC hybrids are located on the isotherms housed inside the
    copper cup shown at the bottom of the figure.
    This figure was published in Nuclear Instruments and Methods A, 
A565, V.M. Abazov et al., pg. 463-537, Copyright Elsevier (2006).
  }
  \label{Cass_OV}
\end{figure}

\begin{figure}
  \begin{center}
    \includegraphics[width=0.95\textwidth]%
    {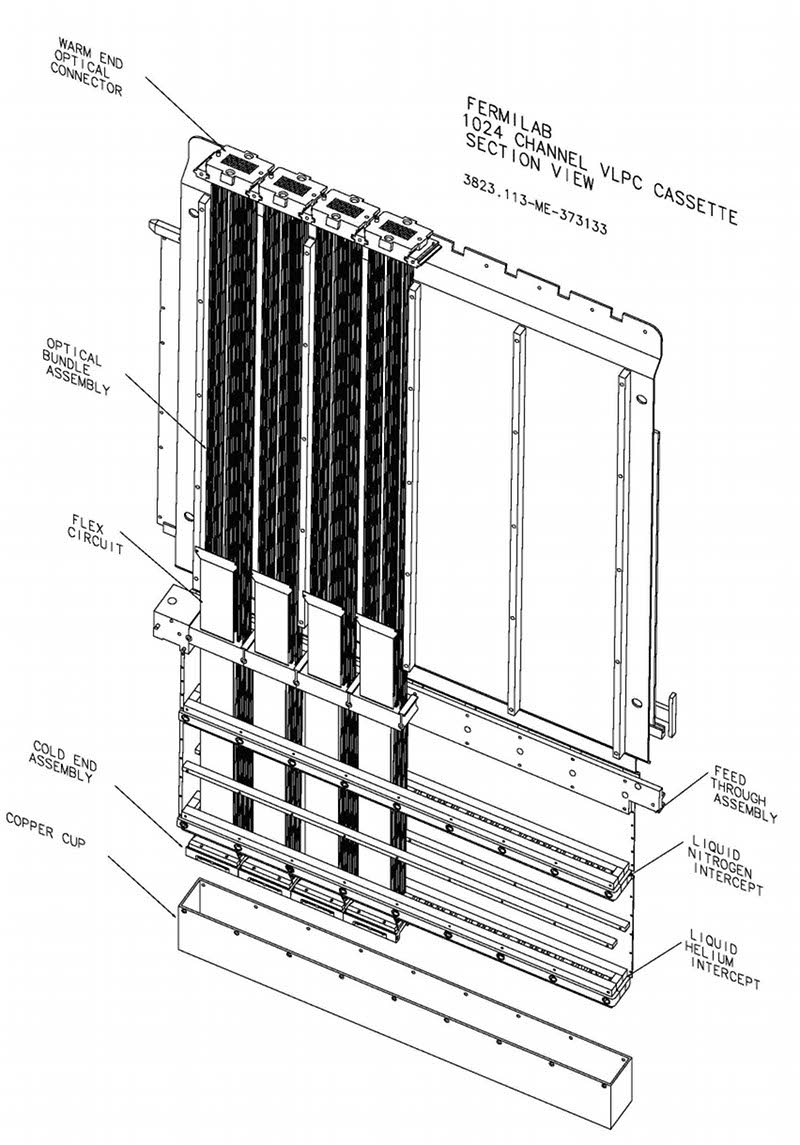}
  \end{center}
  \caption{
    VLPC cassette body with left side body panel and side panels
    removed to show four installed modules.
    This figure was published in Nuclear Instruments and Methods A, 
A565, V.M. Abazov et al., pg. 463-537, Copyright Elsevier (2006).
  }
  \label{Cass_OV_expl}
\end{figure}

The VLPC cassette contains 1024 channels of VLPC readout and is
divided into 8 modules of 128 channels each.
The modules are interchangeable and may be removed for repair.
This is illustrated in figures \ref{Cass_OV} and \ref{Cass_OV_expl}.
Figure \ref{Cass_OV}  shows the full cassette with readout boards
attached. 
Figure \ref{Cass_OV_expl} shows the inner components of the cassette,
with the readout boards and cassette body removed.
Sitting directly over each VLPC pixel is an optical fibre which brings
the light from the detector to the VLPC chip. 
Each cassette module is composed of an optical bundle assembly, a
cold-end electronics assembly, and an assembly of mounted VLPC
hybrids.
The high-cost and delicate nature of the cold-end assembly led to the
development of a design that allows the cold-end assembly to be
removed easily for repair.
Another important design requirement for the cassette regards the
readout electronics.  
Due to the nature of Tevatron operations, it must be possible to
remove and replace the readout electronics boards, the front-end
boards which act as the interface to the data acquisition system,
without removing a cassette from the cryostat.  
The readout electronics are discussed in detail in section
\ref{Sect:Electronics}.

The cassette has a `cold end', that portion of the cassette which lies
within the cryostat, and a `warm end', the portion of the cassette
which emerges from the cryostat and is at room temperature. 
At the cold end, eight `cold-end assemblies', each of 128 channels of
VLPC readout, are hung from the feed-through by the optical bundles
and are surrounded by a copper cup.
Each cold-end assembly consists of sixteen, 8-channel VLPC hybrid
assemblies, the `isotherm' or base upon which they sit, the heater
resistors, a temperature measurement resistor, cold-end flex-circuit
connectors and the required springs, fasteners and hardware (see inset 
in figure \ref{Cass_OV}).   
Running within the cassette body from top to bottom are eight
128-channel optical-bundle assemblies which accept light from the
detector light-guides connected to the warm-end optical connectors at
the top of the cassette and pipe the light to the VLPCs mounted at
the cold end (see figure \ref{Cass_OV_expl}).
The electronic read-out boards are located in rails which are mounted
on the warm end structure and are connected electrically with the
cold-end assemblies via kapton `flex circuits'.
In addition, the electronics boards are connected to a backplane card
and a backplane support structure via multi-pin connectors and
board-mount rails.
The flex circuits and read-out boards are electrically and
mechanically connected by a high density connector assembly. 
For more details on the VLPC cassette design see \cite{D0NIM}.

\begin{figure}
  \begin{center}
    \includegraphics[width=0.75\textwidth]%
    {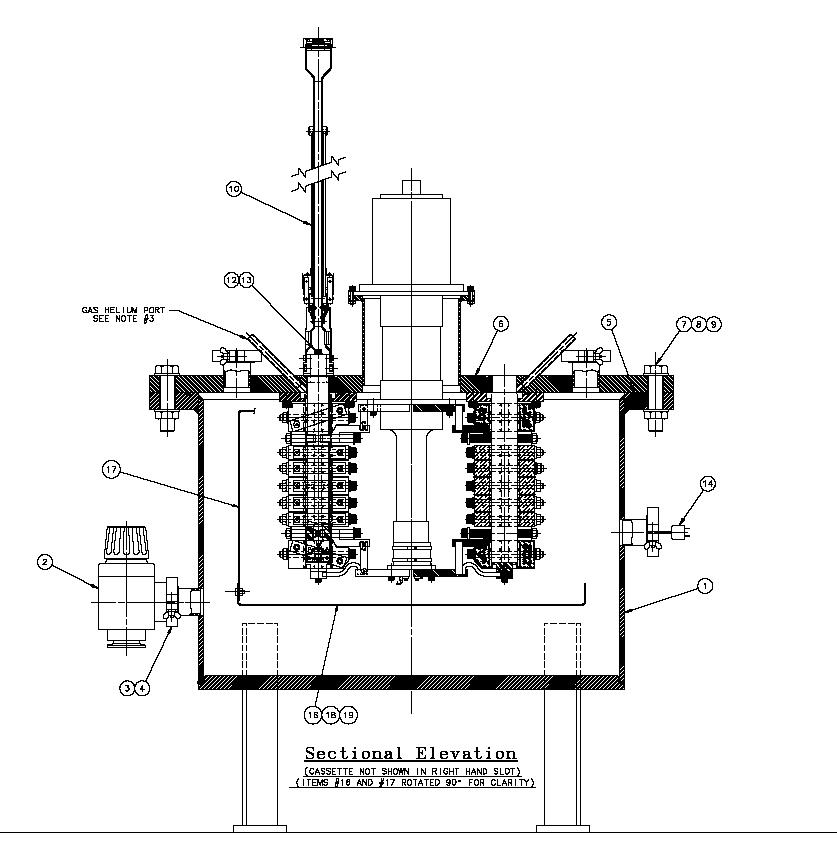}
  \end{center}
  \caption{
    Sectional elevation drawing for the MICE 2-cassette VLPC
    cryostat.
  }
  \label{2CassCryo}
\end{figure}

Since the VLPCs operate at cryogenic temperatures, a helium
cryo-system is required.
A special purpose cryostat has been designed and built for MICE to
allow the VLPCs to be operated at a temperature of $(9.0 \pm 0.1)$~K.
The cryostat has two cassette slots which accept D\O{} 1024 channel VLPC
cassettes (see figure \ref{2CassCryo}). 
The MICE VLPC cryo-systems use Gifford-McMahon cryocoolers (Sumitomo
RDK 415D \cite{Sumi}) to maintain the 9~K operating temperature for
the VLPCs.
This two-stage, commercial cryocooler is located between the two
cassette slots.
The first stage of the cryocooler, operating at 50~K, is used to
provide a heat intercept between the room-temperature cryostat body
and the cold-end assembly.
The cold head of the cryocooler is used to cool the copper isotherm in
order to bring the VLPCs to their operating temperature.  
The 50~Hz mains frequency employed in the UK results in the cryocooler
operating at about 70\% of its rated capacity.
The cryocooler, along with its helium compressor, gas flex lines, and
control cables is provided by the manufacturer as a packaged system. 
The system has worked flawlessly, providing a stable, non-fluctuating
temperature at the cold-end assembly for long operating periods. 

The cassette envelope is a low conductivity, low thermal-expansion box
that seals to the underside of the top lid of the cryostat.  
The insulating vacuum is on the outside of the envelope and the VLPC
cassette, immersed in helium gas, is situated inside the envelope.  
The nominal clearance inside the cassette is made rather tight at
0.6~mm in order to facilitate heat conduction to the cryocooler.
The wall thickness, 0.38 mm, was also reduced to the minimum practical
for fabrication so as to minimise heat conduction along the cassette
envelope from the warm end to the cold end.

\begin{figure}
  \begin{center}
    \includegraphics[width=0.75\textwidth]%
    {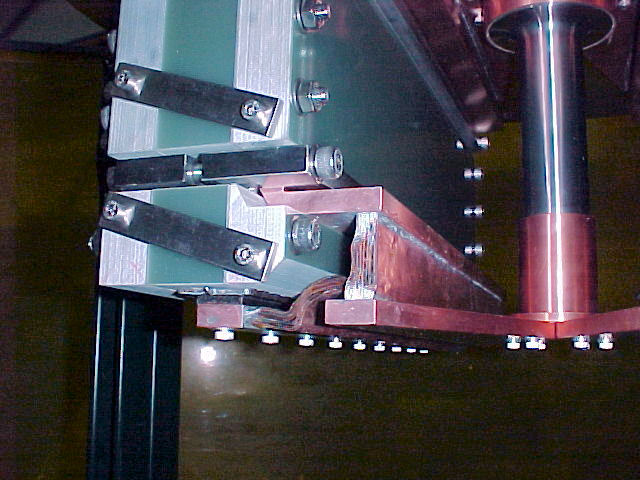}
  \end{center}
  \caption{
    Close up of the lower thermal link flexible section.  
    The cassette envelope may be seen on the left-hand side of the
    picture.
    The copper foils that form the thermal link between
    the cold end and the cryocooler cold head may also be seen.
    The Cernox temperature sensor is mounted to the envelope 
    side of the link.
  }
  \label{lowerlink}
\end{figure}

Thermal links provide the conduction heat-transfer path between the
two stages of the cryocooler and the cassette.  
The thermal links are made from oxygen-free high-purity copper (OFHC),
UNS grade C10100.
The thermal links are 10~mm thick solid copper pieces with a short
flexible segment to accommodate a movement of 2~mm due to thermal
contraction.
The flexible segment is constructed of 35 individual 0.13~mm thick
pieces of high purity copper foil soldered to the solid segments.
The upper thermal link operates at a temperature of around 45--50~K
and has a temperature gradient of 1~K from the cryocooler connection
to the envelope connection.
The lower thermal link (figure \ref{lowerlink}) operates at
temperatures in the range of 6--8~K and has a calculated 0.1~K
temperature drop from the cryocooler to the envelope.

Cernox temperature sensors from Lakeshore Cryogenics \cite{Lakeshore} are mounted on
the cold head of the cryocooler and the lower thermal link near the
cassette envelope. 
A silicon diode temperature sensor from Oxford Instruments is used to
measure the upper link temperature. 
A $100~\Omega$ platinum resistor was also used during the
commissioning tests.  
The sensors are mounted into small copper holders using GE-7031
varnish.
The copper holders themselves are screwed to the surfaces with a
\#4-40 screw with Apiezon `N' grease at the interface.  
Cryogenic quad-lead wire from the sensors is heat-sunk to a small
copper bobbin mounted nearby.
Readout and temperature control is via an Oxford Instruments, ITC503
temperature controller.
A flexible $36~\Omega$ heater element is wrapped around the second
stage of the cryocooler using high purity copper sheeting and hose
clamps. 
The heat transfer surface of the heater is lightly coated with Apiezon
grease.   
Up to 5~W of heat can be supplied by the heater to give gross
temperature control for the cassette cold end.  
Each cassette also has individual heater controls (one heater for each
module in a cassette for a total of 8) to adjust for small, asymmetric
temperature differences.

\begin{figure}
  \begin{center}
    \includegraphics[width=0.90\textwidth]%
    {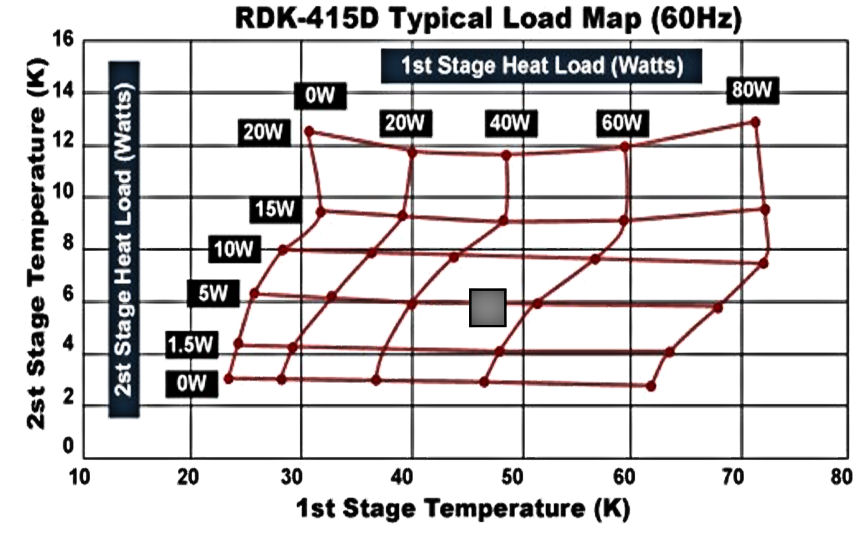}
  \end{center}
  \caption{
    Load map of the Sumitomo RDK 415D cryocooler. 
The square indicates the system operating envelope with no additional
heat (via stage II heater). Figure taken from \cite{Sumi}.  
  }
  \label{loadmap}
\end{figure}

Although we experienced some difficulty in making good
(high-conductance) connections on the thermal links at the cryocooler
head and the cassette envelope, once good link connections had been
made, the system performed extremely well.  
Figure \ref{loadmap} shows the load map for the RDK 415D. 
This agrees well with the operating point expected from heat load
calculations based on the cryostat and VLPC cassette engineering
designs.   
In normal operating conditions, a small amount of heat is added to
stage II using the heater (supplied and controlled by the Oxford
controller) giving better temperature stability.
With full control (Oxford and cassette temperature control on), the
VLPC temperature is controlled at $(9 \pm 0.005)$~K. 

%
\section{Electronics and data acquisition}
\label{Sect:Electronics}

MICE uses the D\O{} Central Fiber Tracker (CFT) optical readout and
electronics system. 
Two Analogue Front End (AFE) boards are mounted on each VLPC cassette.
The AFE boards read out 512 channels each and provide VLPC bias and
temperature control functions.

\subsection{AFEIIt Boards}

\begin{figure}
  \begin{center}
    \includegraphics[width=0.85\textwidth]%
    {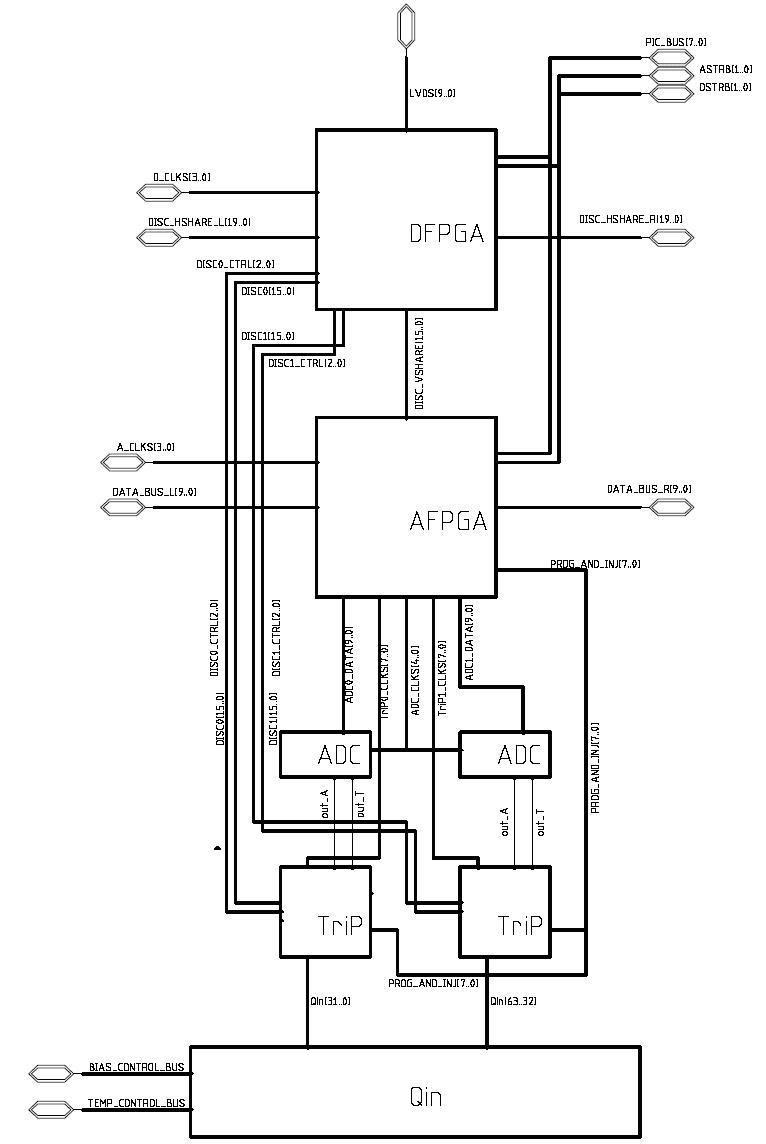}
  \end{center}
  \caption{
    Schematic diagram of the signal path implemented on the AFE-IIt
    board from the TriP-t chips to the VME memory banks.
  } 
  \label{AFESchem}
\end{figure}

The board used in MICE is the second generation front-end 
readout board, the AFE-IIt \cite{AFEIIt}. 
As shown in figure \ref{AFESchem}, the analogue pulses produced by the
VLPCs are input to `Trigger and Pipeline with timing' (TriP-t) chips
on the AFE-IIt boards.
The TriP-t chip generates three outputs for each channel: a digital
discriminator signal; an analogue pulse proportional to the amplitude
of the integrated charge of the input pulse (the A-pulse); and an
analogue pulse (the t-pulse) proportional to the time between the
firing of the discriminator and the closing of the time gate. 

The discriminator output is routed to one of the Field Programmable
Gate Arrays (FPGAs), the DFPGA (��Digital�� or ��Discriminator�� FPGA).
The A-pulse and the t-pulse are stored in 48-sample, analogue
pipelines in the TriP-t chip before being read out upon the receipt of
an external ��Level 1 accept�� (L1ACCEPT) trigger.  
In MICE the discriminator bits, which indicate which channels are
above a predetermined threshold, will be directed to another FPGA
called the AFPGA (��Analogue�� FPGA), and will be used to suppress the
digitization cycle for channels that are below threshold. 
MICE will use a scheme in which only those channels above threshold
are digitised.  
This  scheme will decrease the digitization time for an event. 
The high L1ACCEPT rate (600 kHz) anticipated in MICE has required this
implementation of the Digitise-Readout mode structure. 

In D\O{}, a TriP-t chip receiving a L1ACCEPT signal causes the AFE-IIt
board to direct the TriP-t chip to execute `Acquire', `Digitise', and 
`Readout' cycles for the A-pulse and t-pulse from a fixed location in
the pipeline.
The A-pulse and the t-pulse are then read out over a ribbon cable to
the `SVX Sequencer' (originally designed for initialization, control,
and readout of SVX IIe  silicon vertex detector chips).  
The differing event sizes anticipated in MICE has required a
modification to the Acquire-Digitise-Readout mode structure to be
implemented. 

\subsubsection{AFEIIt Firmware}

For MICE operation, the AFE-IIt firmware is required to accommodate
the expected instantaneous trigger rate of roughly 600 kHz and the
time structure of the muon beam.
The MICE target is designed to dip into the edge of the ISIS beam for
1~ms at the end of the ramp, just before the beam is extracted.
The ISIS proton beam has two 100~ns long bunches that are separated by
200~ns.
At the end of the ramp, when the proton-beam energy is 800~MeV, the
beam takes 324~ns to make one revolution in the synchrotron.
The muon beam for MICE, therefore, appears as a 1~ms wide spill in
which two $\sim 100$~ns long bursts of muons arrive at the experiment
every 324~ns.
The MICE DAQ imposes a trigger constraint such that each L1ACCEPT
trigger produces a trigger hold-off of 550~ns so that triggers cannot
occur on consecutive bursts (separated by 324 ns).  
However, since a burst may contain multiple muons and since data from
an entire burst will be stored in the TriP-t pipeline, a L1ACCEPT can
be associated with multiple muons; the muon rate will be larger than
the trigger rate.  
The timing and spatial separation of multiple muons in a burst will
determine whether they can be reconstructed properly.  
Since this rate is significantly higher than the L1ACCEPT rate in D\O{},
extensive modifications to the AFE-IIt firmware were required,
including modifications to both the digital and analogue data
processing (handling of discriminator bits and of the A-pulse and the
t-pulse respectively).  

\subsection{VLSB Board}

\begin{figure}
  \begin{center}
    \includegraphics[width=0.80\textwidth]%
    {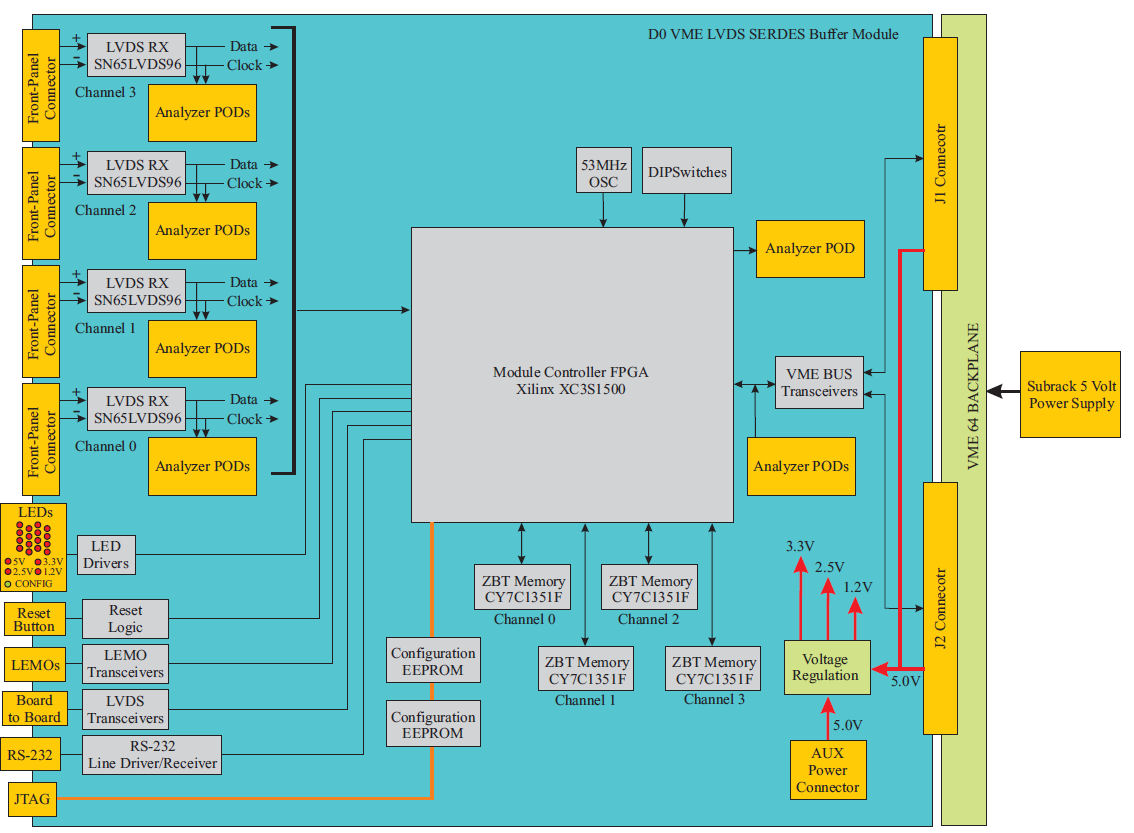}
  \end{center}
  \caption{
    Block diagram of the VLSB board.  
    The diagram shows the module controller, the various blocks of
    memory and the input/output connections.
  }
  \label{VLSB_block}
\end{figure}

The A-pulse and t-pulse are read out over a Low Voltage Differential
Signal (LVDS) path to a custom VME LVDS SERDES
(Serializer/Deserializer) Buffer (VLSB) upon receipt of a L1ACCEPT.
The VLSB module, which was originally designed for D\O{} board testing
and has been adapted for use in MICE, is a VME64 single width, 6U
module. 
The module is a custom LVDS SERDES buffer with 4 LVDS input channels
and can be operated stand-alone.
A VLSB module can receive/generate trigger signals over two Lemo
connectors on the module front panel.
A block diagram of the VLSB card is shown in figure \ref{VLSB_block}
\cite{VLSB}. 

The module can be controlled through two different interfaces: RS232
(front panel) and VME (backplane). 
The VLSB module hosts the `Module Controller FPGA' which handles the
VME and RS232 interfaces, the timing and the diagnostics, and
supervises the data flow operations. 
The FPGA is configured at power-up by two on-board EEPROMs. 
The VLSB module also hosts the four LVDS SERDES receivers used to
convert and de-serialize the input LVDS signals to Low Voltage TTL.
A logic analyzer pod is provided for each LVDS link. 
The Module Controller FPGA manages the input LVDS interface and stores
the data received into the zero-bus turn-around (ZBT) SRAM. 
The FPGA control/status registers and the ZBT SRAM content can be
accessed from both VME and RS232 interfaces.

In MICE, the L1ACCEPT will be formed from signals from Time-of-Flight
counters and signals from the ISIS accelerator; a combination of these
signals will indicate the passage of a through-going muon.
If we assume a 0.5\% occupancy, then there are roughly 30 hits in the
two trackers per muon.
During readout, all 64 DFPGAs will read out a trigger word, a header
word, and 8 words for bitmap data for a total of 10 words per DFPGA. 
Then 30 single hits with 5 words per hit (channel, TriP0 time, TriP1
time, TriP0 charge, and TriP1 charge) yield 150 more words for a total
of $(64 \times 10) + (30 \times 5) = 790$ total words for a MICE
event. 
Assuming 1000 words/event, then for 600 events/spill and one spill
every second, the AFE-IIt boards will have to read out roughly 600,000
words every second to the VLSB buffers via the LVDS links and VME.  
If the data are read out to 16 VLSB modules housed in two crates with
each VME word transfer taking 1~$\mu$s, the 600,000 word readout would
take, reading to the two crates in parallel, roughly 0.3 sec which is 
sufficient for MICE running.

\section{Performance}
\label{Sect:Performance}

\subsection{Cosmic test stand}

The first MICE tracker was installed inside a black carbon fibre
cylinder (referred to as the `coffin') the internal diameter of which
is the same as the warm bore of the spectrometer solenoid.
The patch panel was connected to the coffin via an aluminium flange
and the coffin/patch panel assembly was supported on an aluminium
space frame such that the principal axis of the tracker was vertical.
Trigger scintillators were placed above and below the tracker and a
four-inch layer of lead bricks was placed between the bottom of the
tracker and the lower trigger scintillator to filter out muons with
momentum less than $\sim 210$~MeV/c.  
Data taking began in July 2008 and continued without interruption
until November 2008 when the cosmic stand was disasembled so that it
could be moved to a larger area where both trackers were set up
for further testing. 

\subsection{Reconstruction}

Reconstruction was performed using the standard routines provided
by the G4MICE software package \cite{G4MICE}.
The principal stages in the track reconstruction are:
\begin{itemize}
  \item{\it Cluster finding:} \\
    Following the unpacking of the hits on individual VLPC channels,
    clusters of one or two neighbouring channels from a particular view
    are constructed.
    Due to the readout arrangement, in which seven neighbouring
    scintillating fibres are readout via one VLPC channel, a single
    track can pass through no more than two neighbouring channels. 
    Clusters that have a total reconstructed signal corresponding to
    at least 2.5 photo-electrons (p.e.) are kept to be used in the
    reconstruction of space points;
  \item{\it Space-point reconstruction:} \\
    The space-point reconstruction first searches for `triplets', which
    consist of a cluster in each of the three views on a given
    station. 
    If the internal residual, defined as the perpendicular distance
    from one cluster to the intersection of the other two, is small
    enough, and the light yield of each used cluster high enough,
    the combination of 3 clusters is kept as a space
    point and the clusters are removed from the list of those available.
    All combinations of two clusters that have not already been
    assigned to a point and originate from two different views in the
    same station are then used to create space points refered to as
    `duplets';
  \item{\it Pattern recognition and track fit:} \\
     The pattern recognition code that is used in G4MICE depends on
     the magnetic field in which the tracker is placed.
     The cosmic ray test was performed in the absense of a magnetic
     field and therefore a straight-line track-model is used for the
     pattern recognition and the track fit.

     Sets of 3 points (in different stations) are tested for
     collinearity and, if they pass this test, an extrapolation to the
     remaining two stations is made.
     If a space point falls inside a small road-width around this
     extrapolation, then this point is retained. If more than one
     space point falls inside the road-width for either staion, then the
     nearest is retained.
     Each such combination of 5, 4, or 3 points is passed to the
     Kalman filter track fit \cite{recpack} and the combination with
     the smallest $\chi^{2}$ per degree of freedom is kept. 
     The space points that make up an accepted track are then locked
     so that they can no longer be used in the track search.
     The pattern recognition process repeats until no more tracks are
     found or the list of space points is exhausted.
\end{itemize}

\subsection{Performance}

\begin{figure}
  \begin{center}
    \includegraphics[width=0.95\textwidth]%
    {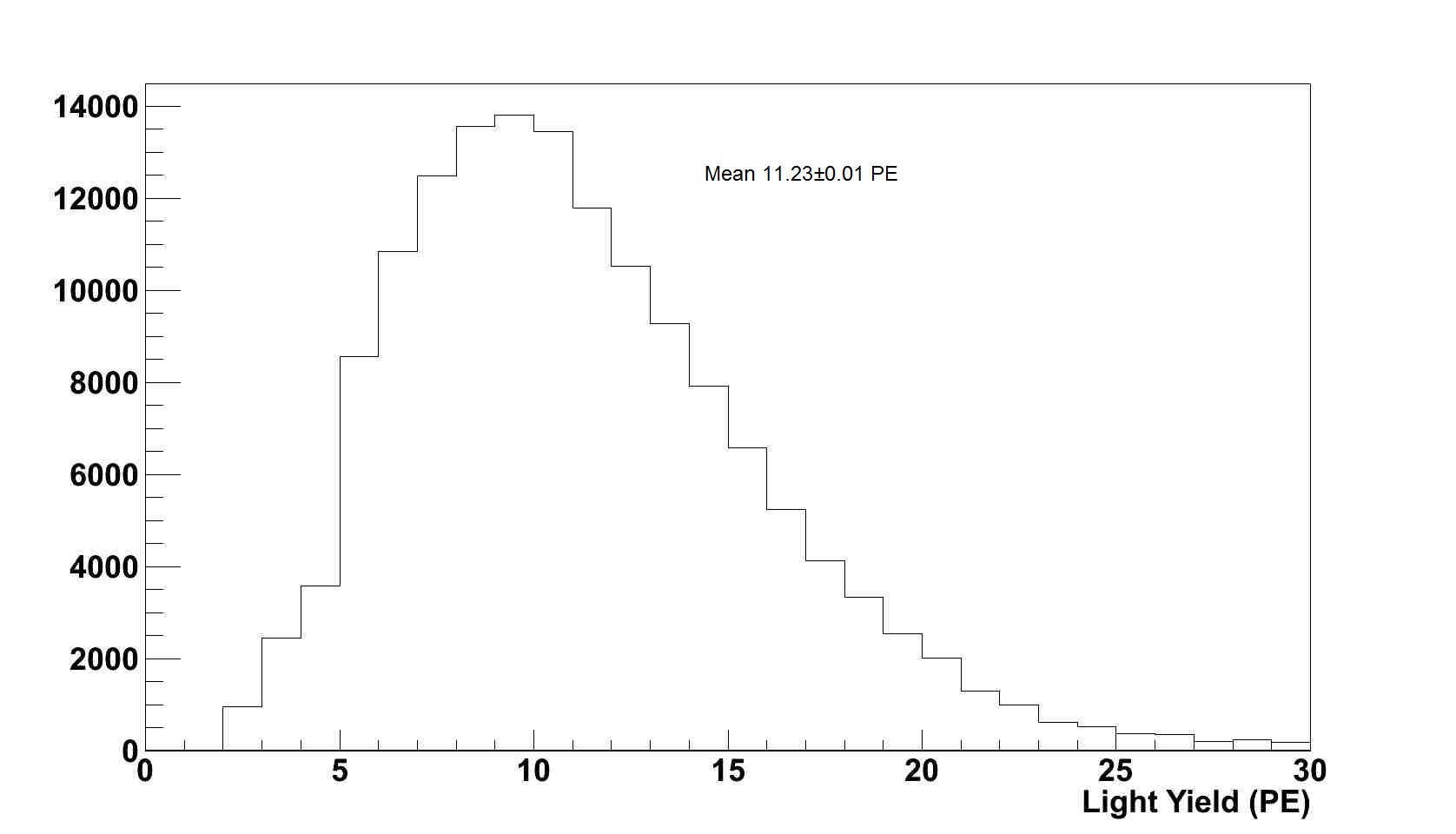}
  \end{center}
  \caption{
Light yield per doublet cluster from cosmic rays, 
corrected for saturation effects in high-gain VLPCs.
  }
  \label{Fig:Performance:LightYield}
\end{figure}

The light-yield, space-point-finding efficiency, and the resolution of
the tracker were determined using points assigned to tracks by the
G4MICE track-finding algorithm described above.
The analogue signals produced by the VLPCs are digitised in an ADC
with an 8-bit dynamic range.
The response of the electronics was calibrated using a light-injection
system.
The one-, two-, three-, and four-photo-electron peaks were clearly
separated from one-another in the calibration data.
In addition, the one-photo-electron peak was clearly separated from
the pedestal.
For each channel, $i$, the calibration data was fitted to extract the
pedestal, $p_i$, and the gain $g_i$.
For channels in which the ADC value (${\rm ADC}_i$) was below the
maximum count of 255, the light yield is given by
$l_i = [{\rm ADC}_i - p_i]/g_i$.
Channels for which ${\rm ADC}_i = 255$ are referred to as `saturated
channels'.
If the ADC value in all channels that make up a cluster falls below
the maximum count, then the cluster light-yield is determined by
summing the $l_i$.
If a cluster contains one or more saturated channels it is not used in
the evaluation of the light yield. To compensate for the saturated channels,
the probability of a channel being rejected in this way is calculated as a function
of the light-yield, and the initial light-yield distribution divided by this distribution of probabilities.
The distribution of the cluster light-yield obtained in this way is shown in
figure \ref{Fig:Performance:LightYield}.
The mean of the cluster light-yield distribution is $11.23 \pm 0.01$
photo electrons (PE) and the most-probable value is
$9.37 \pm 0.03$~PE.

The space-point efficiency was obtained by building tracks using space
points in four of the five stations and extrapolating the track to
the station under test. 
If the extrapolated position fell within the active area of the
station, a search was made for a space point close to the
extrapolated position.
Table \ref{Table:Performance:Summary} reports the space-point efficiency
determined for each station.
The space-point efficiency determined in this way is $99.8 \pm 0.1$\%.
The efficiencies are consistent with what would be expected based on
the measured light yield reported above.

\begin{figure}
  \begin{center}
    \includegraphics[width=0.95\textwidth]%
    {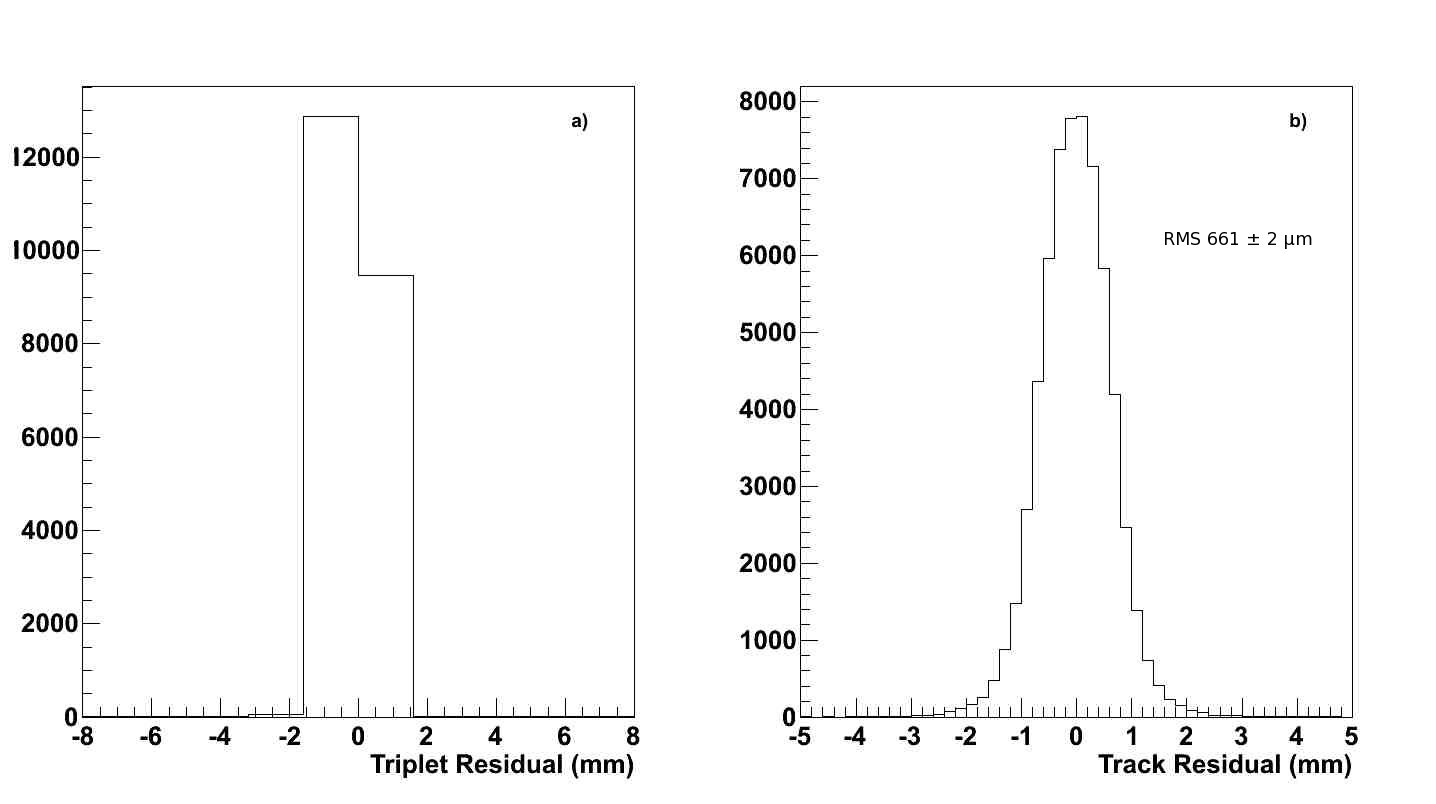}
  \end{center}
  \caption{
    Performance of the MICE tracker evaluated using cosmic rays
    as described in the text.
    a) Triplet residual distribution, consisting of triplets made from single-channel 
    clusters and used in full tracks.
    b) Track residual distribution. The RMS is noted on the figure.
  }
  \label{Fig:Performance:Resolution}
\end{figure}

The intrinsic position resolution of the doublet-layers was evaluated
using triplet space points.
The construction of the stations is such that the fibres in one view
run at an angle of $120^\circ$ with respect to the direction of the fibres in
each of the other two views.
If $u$, $v$, and $w$ are the position of a cluster in each of the
three views that make up the triplet, taking the origin of $u$, $v$, 
and $w$ to be the centre line of the
doublet layer in question, it is expected that the quantity 
$u + v + w = \delta = 0$.
A measure of the intrinsic resolution of a station can therefore be
determined by plotting $\delta$.
Figure \ref{Fig:Performance:Resolution}a shows the distribution of
$\delta$.
The width of the distribution of $\delta$ is well described by the
Monte Carlo and is as expected from the station construction.
To demonstrate that the tracker is precisely aligned and to verify
the treatment of material in the track fit, the resolution
was determined using reconstructed tracks. 
Tracks composed of five space points were re-fitted, removing one
space point at a time. 
Each of the five new four-space-point tracks was extrapolated to the
station under test and the perpendicular distance (residual) between the
extrapolated position and each of the channels forming the removed space point calculated.
The distributions of the total residual is shown in figure
\ref{Fig:Performance:Resolution}b.
The RMS of the residual distributions is $661 \pm 2~\mu$m as expected given
a channel resolution of $470 \mu$m and multiple Coulomb scattering in the doublet layers.
\begin{table}
  \begin{center}
    \begin{tabular}{cc}
      \hline
      Station & Efficiency $(\%)$  \\
      \hline
      \hline
      1       & $99.8 \pm 0.1$   \\
      2       & $99.9 \pm 0.1$   \\
      3       & $99.7 \pm 0.1$   \\
      4       & $99.9 \pm 0.1$  \\
      5       & $99.8 \pm 0.1$   \\
      \hline
    \end{tabular}
  \end{center}
  \caption{
    Efficiency of the five stations that make up the scintillating
    fibre tracker determined using cosmic rays.
  }
\label{Table:Performance:Summary}
\end{table}

%
\section{Summary}
\label{Sect:Summary}

The design of the scintillating-fibre trackers for the international
Muon Ionisation Cooling Experiment (MICE) which is under construction
at the Rutherford Appleton Laboratory has been described.
The construction techniques, including the quality assurance
procedures, and the optical readout system have been described in
detail.
Finally, the performance of the first completed device has been
presented and shown to meet the design specifications.
The two trackers for MICE are presently being commissioned in a
cosmic-ray test stand and will be installed in the MICE spectrometer
magnets in the near future. 

%
%
\section*{Acknowledgements}

The final design of the MICE tracker was developed from the concept
proposed by P.~Gruber and E.~McKigney in
\cite{GruberMcKigney2001CERN}\cite{GruberMcKigney2001ICL}
and P.~Janot in \cite{Janot2001CERN}.
We are grateful to the D\O{} collaboration for the loan of a number of 
VLPC cassettes and for their support and advice throughout the MICE
tracker project. 
We are indebted to the MICE collaboration, which has provided the
motivation for, and the context in which, the work reported here was
carried out. We would like to acknowledge AC Precision, Wantage,
Oxfordshire, UK for help and advice during the manufacture of the
optical connectors and G-TECH, Tsukuba, Japan for advice during the
manufacture of the light-guides. 
We would also like to acknowledge the hospitality of FNAL, KEK, and
RAL where test exposures were carried out and the various institutes
around the world (FNAL, Imperial, KEK, Osaka, and RAL) at which
tracker workshops have been held. 

This work was supported by the Science and Technology Facilities Council
under grant numbers
PP/E003214/1, PP/E000479/1, PP/E000509/1, PP/E000444/1, and through SLAs
with STFC-supported laboratories.
This work was also supportedby the Fermi National Accelerator Laboratory,
which is operated by the Fermi Research
Alliance, under contract No. DE-AC02-76CH03000 with the U.S. Department of
Energy, and
by the U.S. National Science Foundation under grants
PHY-0301737,PHY-0521313, PHY-0758173 and PHY-0630052.
The authors also acknowledge the support of the World Premier International
Research Center Initiative (WPI Initiative), MEXT, Japan.

%
\clearpage
\bibliographystyle{styles/utcaps}
\bibliography{TrackerPaper}
%
\end{document}